\documentclass[usenatbib]{mnras}

\usepackage{hyperref}
\usepackage{lmodern}
\usepackage{amsmath,amssymb}
\usepackage[normalem]{ulem}
\usepackage{textcomp}
\usepackage{bm}
\usepackage{graphicx}
\usepackage{psfrag}
\usepackage{aas_macros}
\usepackage[usenames,dvipsnames]{xcolor}
\usepackage[utf8]{inputenc}
\usepackage{lipsum} 
\usepackage{multirow}
\usepackage{lineno}

\newcommand{\mlow}{\ensuremath{M < 40}}
\newcommand{\mmid}{\ensuremath{40 \leq M \leq 80}}
\newcommand{\mhigh}{\ensuremath{M > 80}}
\newcommand{\mdhigh}{MD\textsubscript{high}}
\newcommand{\mdlow}{MD\textsubscript{low}}

\definecolor{darkgreen}{rgb}{0,0.5,0}

\hypersetup{
    bookmarks=true,         
    unicode=false,          
    pdftoolbar=true,        
    pdfmenubar=true,        
    pdffitwindow=false,     
    pdfstartview={FitH},    
	pdftitle={Main},
    pdfauthor={Divyajyoti et al.},
    pdfsubject={Gravitational Waves},   
    pdfkeywords={gravitational waves, eccentric compact binaries, general relativity},
    pdfnewwindow=true,      
    colorlinks=true,       
    linkcolor=red,          
    citecolor=Blue,        
    filecolor=magenta,      
    urlcolor=Blue,           
    linktocpage=true
} 

\allowdisplaybreaks[4]

\DeclareFontFamily{OT1}{pzc}{}
\DeclareFontShape{OT1}{pzc}{m}{it}{<-> s * [1.10] pzcmi7t}{}
\DeclareMathAlphabet{\mathpzc}{OT1}{pzc}{m}{it}

\pubyear{\the\year{}}

\begin{document}
\label{firstpage}
\pagerange{\pageref{firstpage}--\pageref{lastpage}}

\title{Mapping the star formation peak with LIGO A\# and Next-Generation detectors}

\author[Divyajyoti et al.]{
Divyajyoti\thanks{E-mail: divyajyoti.physics@gmail.com},
Stephen Fairhurst,
Mark Hannam,
and Mukesh Kumar Singh
  \\
  Gravity Exploration Institute, School of Physics and Astronomy, Cardiff University, Cardiff, CF24 3AA, United Kingdom}

\maketitle

\begin{abstract}
Measuring the redshift evolution of star formation rate density is crucial in understanding the origin and evolution of galaxies and large scale structure in the universe. It is currently measured with electromagnetic probes, however, these probes often track luminosity, which is then converted to star formation rate (SFR) depending on various factors such as initial mass function, dust extinction, etc. Gravitational waves provide an independent method to constrain SFR at high redshifts by tracking the redshift evolution obtained from analysis of binary black hole mergers. In this study we explore three population models for star-formation combined with an \textit{inverse} time-delay model and demonstrate that it is possible to obtain bounds on the peak of redshift distribution with a network of upgraded LIGO detectors (such as LIGO-A\#). For a year of observation, using simulated signals with a merger rate peak at $z_\text{peak}=1.5$, a network of LIGO detectors at A\# sensitivity is able to constrain the peak of merger rate with a precision of $\pm 0.1$. Further, we obtain the results with a next-generation network (of Cosmic Explorer and Einstein Telescope) and conclude that the redshift distribution will be extremely well measured, with a precision of $\pm 0.02$, with future detectors.
\end{abstract}

\begin{keywords} 
gravitational waves -- star formation -- binary black hole population
\end{keywords}

\date{\today}

\section{Introduction}
\label{sec:intro}

The redshift evolution of the binary black hole (BBH) merger rate encodes key information about the mechanisms driving their formation and coalescence, including the astrophysical environments in which they originate, the underlying star formation rate (SFR), and the distribution of time delays between formation and merger~\citep{Vitale:2018yhm, Mapelli:2017hqk, Santoliquido:2020axb, Sgalletta:2024ozw}. Putting constraints on the peak of BBH mergers and SFR is fundamental in understanding the origin and evolution of galaxies, and, when combined with data from galaxy catalogs, can provide estimates on metallicity in stars, heavy elements in the galaxies, and probe the formation of large-scale structure. 

SFR is currently measured using various electromagnetic (EM) probes \citep{Madau:2014bja, Behroozi:2012iw, Botticella:2011nd, 2025A&A...701A.152J}, but these may lead to overestimates, particularly at high redshift during the peak era of galaxy growth. For UV-based measurements~\citep{Madau:2014bja, Dave:2007km, 2025ApJ...994...14P, 2026MNRAS.545f1995B, Dave:2011ak}, a likely culprit is the luminosity-weighted dust corrections, which may be too large, although it is often asserted that UV data tend to underestimate SFRs in very dusty, luminous galaxies. On the other hand, IR-based SFRs may be overestimated, as they were for some high-redshift galaxies in earlier \textit{Spitzer}~\citep{Werner:2004zk} studies. While, on average, the SFRs inferred for individual galaxies may be correct, the luminosity function extrapolations could be too large~\citep{Madau:2014bja}. Although, in recent years, observations from the Hubble Space Telescope and the James Webb Telescope have expanded the catalogues, prompting better constraints of the star formation rate density (SFRD) at higher redshifts~\citep{Perez-Gonzalez:2025bqr, Finkelstein:2015fny, 2026MNRAS.546f2267C}. As far as the EM observations and data are concerned, while the peak era of cosmic star formation has been extensively mapped, it remains difficult to accurately pinpoint the redshift of maximum SFRD within a range of $\Delta z = 1$. The fitting function in \cite{Madau:2014bja} places this peak at $z \approx 1.85$ while a more recent estimate made by \cite{Moyses:2026pqu} puts the peak at $z=2.6$ by jointly analysing SFRD with recent data from baryon acoustic oscillations and type Ia supernovae. Uncertainties in the faint-end slope of the IR luminosity function and in extinction corrections for the UV luminosity function still dominate at this peak era of cosmic star formation~\citep{Madau:2014bja}. At $z > 4$, galaxy surveys have been strongly biased toward UV-bright galaxies and may underestimate both SFR densities and stellar mass densities~\citep{Madau:2014bja}.

The most recent catalog of gravitational wave (GW) events ~\citep{LIGOScientific:2026sit, LIGOScientific:2026ctl} released by the LIGO-Virgo-KAGRA (LVK) collaboration reports compact binary mergers from over 300 sources, which include BBH mergers, binary neutron star mergers, and neutron-star--black-hole mergers. With the current sensitivities~\citep{Capote:2024rmo, LIGO:2024kkz, Virgo:2019juy} of the LIGO \citep{LIGOScientific:2014pky}, Virgo \citep{Acernese:2015gua}, and KAGRA \citep{Akutsu:2020his} detectors, \cite{LIGOScientific:2025pvj} infer that 99\% of detectable BBHs fall below a redshift of $z=1.5$. With the current GW events, it has been observed that the BBH merger rate increases with redshift as a power law $(1+z)^\kappa$ with $\kappa=2.5^{+0.7}_{-0.7}$~\citep{LIGOScientific:2026ctl}. 
While this is consistent with the observed evolution of SFRD in $z\lesssim1$, it is limited by the reach of the current GW network, where detections at $z>1$ are few and with large error bars on redshift estimates. Although \cite{Bers:2025tei} recently showed that a templated background search method can provide constraints on the shape of the redshift distribution beyond the peak of star formation, the constraints have large error bars.

As improvements in sensitivity continue to be made for the forthcoming observing runs of the LVK detectors, there are proposals for major upgrades of LIGO detectors to LIGO-A\# sensitivity~\citep{McClelland:T1500290-v3, post_o5_report}. Furthermore, the next-generation (XG) ground-based GW detectors, Cosmic Explorer (CE) \citep{McClelland:T1500290-v3, Dwyer:2014fpa, LIGOScientific:2016wof, Evans:2021gyd, Evans:2023euw} and Einstein Telescope (ET) \citep{Punturo:2010zz, Hild:2010id, ET:2025xjr}, are proposed to have a strain sensitivity that is an order of magnitude better than the current detectors and a low-frequency cutoff as low as $5$\,Hz. With these sensitivity increases, the reach of ground-based GW detector network is expected to expand with access to compact binary sources up to $z \approx 4$ for A\# and $z>20$ for XG~\citep{post_o5_report, Evans:2023euw}. With access to binaries at higher redshifts, it will be possible to put constraints on the SFR and obtain the overall redshift distribution for compact binary sources. This can further shed light on many of the fundamental questions, such as whether there is a characteristic cosmic epoch of the formation of stars and heavy elements in the galaxies, or how the cosmic history of star formation compares with the history of mass-accretion-only massive black holes as traced by luminous quasars. Measuring the SFR can also shed light on galaxy evolution and allow us to trace their history, probe the large-scale structure, and provide insight into the role of dark matter~\citep{Moster:2012fv, Perez-Gonzalez:2025bqr}.

Moreover, the inferred redshift distribution can provide important clues regarding BBH formation channels~\citep{Fishbach:2018edt}. For instance, a redshift evolution that follows the SFR convolved with a distribution of time delays between formation and merger, points to an isolated binary evolution channel~\citep{Dominik:2014yma, Belczynski:2016obo}. On the other hand, in the dynamical formation channel, the evolution of BBH mergers is influenced by the formation rate of dense star clusters and the merger timescale of binaries within them~\citep{Chatterjee:2016thb, Rodriguez:2017pec}. Other formation scenarios, such as primordial BBHs, which are expected to largely follow the dark matter distribution~\citep{Mandic:2016lcn, Koushiappas:2017kqm}, or gravitational leakage~\citep{Dvali:2000hr, Deffayet:2007kf, Pardo:2018ipy}, or strongly lensed BBH systems~\citep{Broadhurst:2018saj, Smith:2017jdz}, would lead to different features in the inferred redshift distribution. 

Finally, metallicity-dependent cosmic star formation history can be derived by combining statistical galaxy properties constrained by galaxy surveys~\citep{Boco:2020pgp, 2019MNRAS.488.5300C, 2021MNRAS.508.4994C, Dominik:2013tma, Boco:2019teq} or extracted from cosmological simulations of galaxy evolution~\citep{Schaye:2014tpa, Crain:2015poa, Pillepich:2017fcc, Nelson:2018uso, Pakmor:2022vnb}. Constraints obtained from GW data will provide a new, complementary way to study cosmic star formation, the early (iron) enrichment history, and the properties of low-mass galaxies in the reionisation epoch~\citep{Chruslinska:2022ovf, Vitale:2018yhm, Turbang:2023tjk, Fishbach:2023pqs, Singh:2023cxn}. Recent works \citep[][for instance]{Vitale:2018yhm, Mapelli:2017hqk, Santoliquido:2020axb, Sgalletta:2024ozw} explore the effects of metallicity fraction, time-delay models, and common envelope on merger rates and star formation.

In this work we analyse a year's worth of BBH population considering three population models for the redshift distribution: Madau-Dickinson (MD) with inverse time-delay (peak of redshift distribution at $z\sim1.5$), and two modified Madau-Dickinson models, one with peak of redshift at $z\sim1.2$ (referred to as \mdlow) and the other with peak of redshift distribution at $z\sim2$ (referred to as MD\textsubscript{high}). We perform population inference on these simulated BBH mergers to get constraints on peak of the merger rate, $z_\text{peak}$, and ultimately reconstruct the astrophysical distribution of BBH merger rates in the range of $0 < z < 10$. We find that not only is it possible to constrain the peak of merger with XG network, but even with a network of LIGO detectors at A\# sensitivity, we are able to constrain the merger peak at $z_\text{peak} = 1.53^{+0.12}_{-0.10}$ for MD population model. Of course, once the XG network is online, the redshift peak and the overall reconstructed curve are much better constrained ($z_\text{peak} = 1.49_{-0.02}^{+0.02}$)\footnote{The numbers quoted for XG are projected numbers for a year's worth of detections calculated by rescaling the bounds obtained by analyzing 8 weeks of data. Further details can be found in Sec.~\ref{subsec:res:pop-inf}.}. We obtain similar bounds for \mdhigh\ and \mdlow\ population models. 

Further, to provide an insight into which masses of BBH are contributing most to our ability to infer the redshift of the peak merger rate, we divide the population into three mass ranges and obtain independent results for each range.  While the population is skewed to lower mass BBH, the more massive mergers emit a stronger GW signal.  We assess how these competing effects combine to impact our ability to infer the redshift distribution of the signals.

This paper is organized as follows: Sec.~\ref{sec:method} outlines the details of our analysis methods, population models, and detector networks, in Sec.~\ref{sec:results} we present the results, and we conclude in Sec.~\ref{sec:concl}.


\section{Methodology}
\label{sec:method}

We simulate a year's worth of binary black hole signals, using a particular star formation model based on \cite{Madau:2014bja} and a time delay between formation and merger distribution inversely proportional to the delay. We choose a local merger rate of 22 Gpc\textsuperscript{-3} yr\textsuperscript{-1}~\citep{KAGRA:2021duu}. From this population, we identify the signals that would be observed in a given detector network and estimate parameter uncertainties using a Fisher Matrix analysis~\citep{Dupletsa:2022scg}. From the Fisher errors and covariance matrices, we obtain posteriors for the simulated BBH events and use these as the input to a population inference analysis~\citep{Talbot:2024yqw} to infer the underlying astrophysical distribution of merger rate against redshift. 

We perform the analysis for two different 3-detector networks with LIGO A\# and next-generation (XG) sensitivities, and we compare the results between these networks to quantify their capabilities in constraining the shape and peak of the star-formation distribution. In the following sub-sections we elaborate on each of these steps in detail and include the specific population models, detectability criteria, and PE settings used in our analyses. We repeat the analysis with several different true values for the redshift of peak star formation, and for different mass ranges of BBH to investigate the sensitivity of the results to the underlying astrophysics.  


\subsection{Detector networks}
We use the LIGO-Livingston (L), LIGO-Hanford (H), and LIGO-India (I)~\citep{LI:proposal, Saleem:2021iwi, Unnikrishnan:2023uou, Pandey:2024mlo} detectors to form the A\# network, where the PSD curve for all three detectors is taken to be A\# strain sensitivity~\citep{ASharp_sensitivity}. We use the same network of detectors with A+ sensitivity~\citep{O5c_sensitivity} as well for a preliminary SNR comparison shown in Fig.~\ref{fig:input-pop-z}. For both of these networks, the lower cutoff frequency is taken to be 10~Hz. For the next-generation network (XG), we choose a 40 km Cosmic Explorer (CE40) at the default sensitivity given in \cite{CE_sensitivity}, a 20 km Cosmic Explorer (CE20) at the default sensitivity for CE20 in \cite{CE_sensitivity}, both located in the US, and an Einstein Telescope (ET) in triangle configuration at ET10 sensitivity~\citep{ET_sensitivity_CoBA} used in the CoBA study~\citep{ET:2025xjr} located in Italy (for convenience we've chosen the same location as the current Virgo detector). For XG, we take the lower frequency cut-off as 5~Hz. The location and orientation for all detectors are given Appendix~\ref{app:detectors}.


\subsection{Population models}
\label{subsec:method:setup}

\begin{table}
\centering
\def\arraystretch{1.3}
\begin{tabular}{|c|c|c|}
\hline
\textbf{Parameter} & \textbf{Prior} & \textbf{Range} \\ \hline
$m_1$, $m_2$ & \textsc{Power-Law + Peak} & $2 \text{ - } 100$ \\ \hline
$q=m_1/m_2$ & \textsc{Power-Law + Peak} & $1 \text{ - } 50$ \\ \hline
$z$ & \begin{tabular}[c]{@{}c@{}}\textsc{Madau-Dickinson}\\ with time-delay\end{tabular} & $0 \text{ - } 10$ \\ \hline
$a_1$, $a_2$ & Uniform & $0 \text{ - } 0.99$ \\ \hline
$\theta_1$, $\theta_2$ & Uniform sine & $0 \text{ - } \pi$ \\ \hline
$\phi_{12}$ & Uniform & $0 \text{ - } 2\pi$ \\ \hline
$\phi_\text{JL}$ & Uniform & $0 \text{ - } 2\pi$ \\ \hline
$\theta_\text{JN}$ & Uniform sine & $0 \text{ - } \pi$ \\ \hline
$\phi_c$ & Uniform & $0 \text{ - } 2\pi$ \\ \hline
$\alpha$ & Uniform & $0 \text{ - } 2\pi$ \\ \hline
$\delta$ & Uniform cos & $-\pi/2 \text{ - } \pi/2$ \\ \hline
$\psi$ & Uniform & $0 \text{ - } \pi$ \\ \hline
\end{tabular}
\caption{Probability distribution models and ranges for parameters used for simulating binaries.}
\label{table:pop_models_extrinsic}
\end{table}

\begin{figure*}
    \centering
    \includegraphics[trim=0 0 0 20, clip, width=0.495\linewidth]{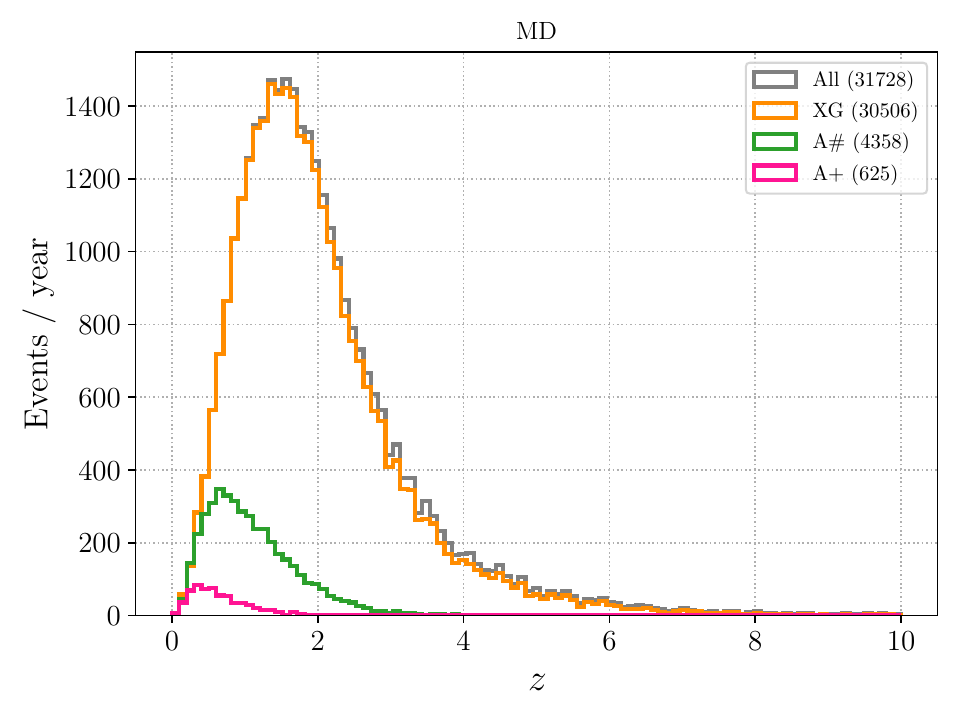}
    \includegraphics[width=0.47\linewidth]{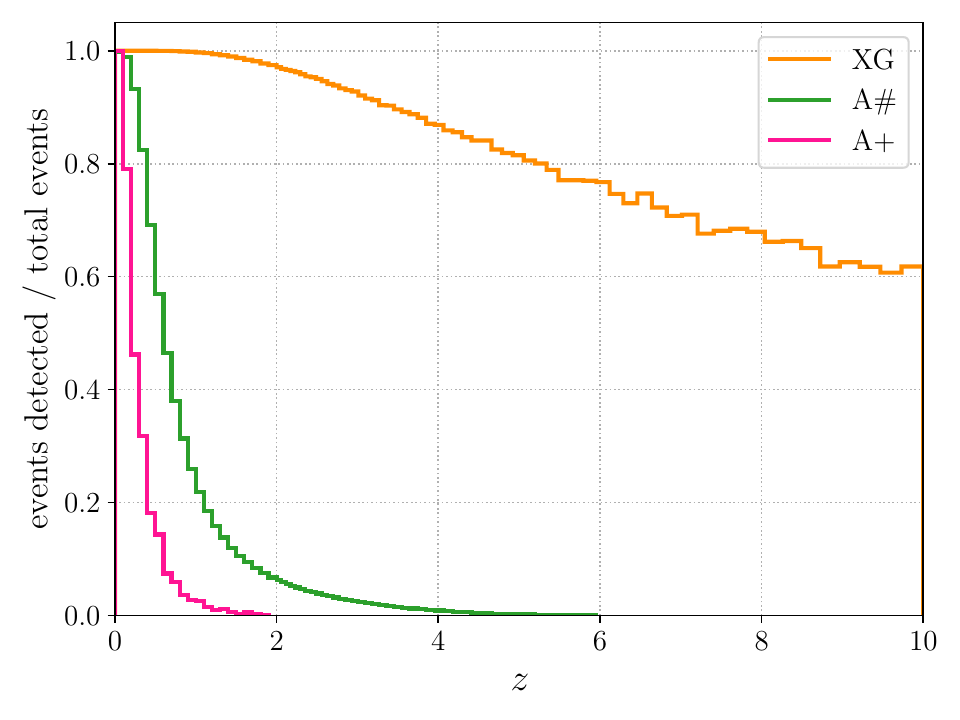}
    \caption{Left: Input population for MD population model. The histogram in grey shows all the simulated BBH systems for 1 year of data up to redshift of 10. The detected events for XG and A\# networks have been plotted in orange and green histograms, respectively. For reference, the events observable with A+     sensitivity are shown in pink. The numbers given in the parentheses are the total number of events for `All' and the detected events for respective networks. We require use a network SNR $\ge 12$ for an event to be detected (as discussed in detail in \ref{subsec:method:fm}). Right: Efficiency curves for XG, A\#, and A+ the networks as a function of redshift. The variations seen in the efficiency curve for high redshift values can be attributed to small-number statistics.
    }
    \label{fig:input-pop-z}
\end{figure*}

\begin{table*}
\def\arraystretch{1.3}
\setlength{\tabcolsep}{12pt}
\begin{tabular}{|c|c|cc|cc|}
\hline
\multirow{2}{*}{\begin{tabular}[c]{@{}c@{}}Population \\ model\end{tabular}} &
  \multirow{2}{*}{\begin{tabular}[c]{@{}c@{}}Total no. \\ of sources\end{tabular}} &
  \multicolumn{2}{c|}{Detection percentage} &
  \multicolumn{2}{c|}{\begin{tabular}[c]{@{}c@{}}No. of sources analysed\\ (equivalent observation time)\end{tabular}} \\ \cline{3-6} 
 &
   &
  \multicolumn{1}{c|}{A\#} &
  XG &
  \multicolumn{1}{c|}{A\#} &
  XG \\ \hline
\begin{tabular}[c]{@{}c@{}}\mdlow\\ ($z_\text{peak} \sim 1.20$)\end{tabular} &
  22,760 &
  \multicolumn{1}{c|}{16.6\%} &
  96.4\% &
  \multicolumn{1}{c|}{\begin{tabular}[c]{@{}c@{}}3667\\ (1 year)\end{tabular}} &
  \begin{tabular}[c]{@{}c@{}}3931\\ ($\sim$ 9 weeks)\end{tabular} \\ \hline
\begin{tabular}[c]{@{}c@{}}MD\\ ($z_\text{peak} \sim 1.50$)\end{tabular} &
  31,728 &
  \multicolumn{1}{c|}{13.4\%} &
  95.7\% &
  \multicolumn{1}{c|}{\begin{tabular}[c]{@{}c@{}}4110\\ (1 year)\end{tabular}} &
  \begin{tabular}[c]{@{}c@{}}4402\\ ($\sim$ 8 weeks)\end{tabular} \\ \hline
\begin{tabular}[c]{@{}c@{}}\mdhigh\\ ($z_\text{peak} \sim 2.01$)\end{tabular} &
  49,086 &
  \multicolumn{1}{c|}{10\%} &
  94\% &
  \multicolumn{1}{c|}{\begin{tabular}[c]{@{}c@{}}4704\\ (1 year)\end{tabular}} &
  \begin{tabular}[c]{@{}c@{}}4861\\ ($\sim$ 5 weeks)\end{tabular} \\ \hline
\end{tabular}
\caption{Total number of sources generated in each population model for 1 year of BBH population taking local merger rate as 22 Gpc$^{-3}$ yr$^{-1}$, detection percentage of each model considering the  $\rho^{\rm{opt}} \geq 12$, and the number of sources analysed for population inference for A\# and XG detector networks.}
\label{tab:num-sources}
\end{table*}

We simulate a year's worth of binary black hole population using the population models described below with the help of the python package \texttt{gwforge}~\citep{Chandra:2024dhf}. For the mass distribution, we consider the \textsc{Power-Law + Peak} model outlined in \cite{KAGRA:2021duu}. The primary mass distribution is given by:
\begin{equation}
p(m_1) = \big[(1-\lambda_{\rm peak})\mathcal{B}(m_1) + \lambda_{\rm peak}G(m_1)\big]S(m_1|m_\text{min}, \delta_m),
\end{equation}
where
\begin{equation}
\begin{aligned}
\mathcal{B}(m) &= \mathcal{C}m^{-\alpha},\ \ m < m_{max},\text{\ \ \ \ and}\\
G(m) &= \frac{1}{\sqrt{2\pi}\sigma_m}\big[e^{-(m-\mu_m)^2/2\sigma_m^2}\big].
\end{aligned}
\end{equation}
Here $\mathcal{C}$ is a normalization constant, and $S(m|m_\text{min}, \delta_m)$ is the smoothing function given by Eq.~(B5) of \cite{KAGRA:2021duu}. The values of population parameters used for the distribution are: $\alpha=3.37,\ \delta_m=5.23,\  m_\text{min}=4.89,\ m_\text{max}=88.81,\ \lambda_\text{peak}=0.04,\ \mu_m=33.60,\ \sigma_m=4.59$ taken from GWTC-3 population \citep{KAGRA:2021duu}.\footnote{Although we use the mass parameters from the GWTC-3 catalog, the overall distribution is comparable with GWTC-5, and we don't expect this choice to have a major impact on our results.} The mass ratio distribution is given by a power law that also includes the smoothing term and is given as
\begin{equation}
    p(q) = q^\beta S(m_1 q|m_\text{min}, \delta_m)
\end{equation}
with $\beta=0.76$. We have distributed the component masses ($m_i$) in the range of [2, 100]~$M_\odot$ and the mass ratio ($q=m_1/m_2$) in the range of [1, 50]. 

The spins of the binaries are distributed using an isotropic distribution. The probability density models for the spin magnitudes ($a_i$) and spin angles ($\theta_i$, $\phi_{12}$, $\phi_\text{JL}$) are given in Table \ref{table:pop_models_extrinsic},\footnote{More information on the definition of the spin parameters \texttt{a\_1, a\_2, tilt\_1, tilt\_2, phi\_12} and \texttt{phi\_jl} can be found at \cite{pesummary}.} where we have also listed the models and parameter ranges used for inclination angle ($\theta_\text{JN}$), sky location angles ($\alpha, \delta, \Psi$), and coalescence phase ($\phi_c$) used for simulating the binaries.

To generate an appropriate redshift distribution accessible to an XG detector network, we have employed the Madau-Dickinson star formation model~\citep{Madau:2014bja} with an inverse time-delay distribution~\citep{Dominik:2012kk, Dominik:2013tma} on the redshift. The star formation rate for Madau-Dickinson model can be written as
\begin{equation}
    \psi(z|\gamma, \kappa, z_p) = 0.015 \frac{(1+z)^{\gamma}}{1+\big(\frac{1+z}{1+z_p}\big)^\kappa}
    \label{eq:madau-dickinson}
\end{equation}
with $\gamma = 2.7$, $\kappa=5.6$, and $z_p=1.9$. The time delay distribution~\citep{Fishbach:2018edt} given by 
\begin{equation}
    p(\tau) \propto \frac{1}{\tau}, \quad t_{\mathrm{min}} < \tau < t_{\mathrm{max}},
\end{equation}
is the probability density for the time delay $\tau$ between formation and merger of the system.  While simulating the population, we have taken $t_\text{min}=0.02$~Gyr, in accordance with \cite{Regimbau:2012ir}, and $t_\text{max}$ as the age of the universe. Convolving the star formation rate with the time-delay distribution, we can write the volumetric total merger rate at redshift $z_m$ in source frame as~\citep{Vitale:2018yhm}\footnote{Here the subscript `$f$' is used to denote quantities at formation and `$m$' is used for merger.},
\begin{equation}\label{eq:R_z}
    \mathcal{R}_m(z_m) \propto \int_{z_m}^\infty dz_f \frac{dt_f}{dz_f} \psi(z_f) p(t_m - t_f)
    \, .
\end{equation}

In the remainder of this paper, we fit the merger rate given by Eq.~\eqref{eq:R_z} using the same functional form as used for the Madau-Dickinson star formation rate in Eq.~\eqref{eq:madau-dickinson}.  This approximation is shown to be an excellent fit to the merger rate in Appendix \ref{app:fit_parameters}, and in particular Fig.~\ref{fig:fitted-curve}. Thus, we write
\begin{equation}
 \label{eq:R_z_fit}
    \mathcal{R}_m(z) \propto 
    \frac{(1+z)^{\gamma'}}{1+\left(\frac{1+z}{1+z'_p}\right)^{\kappa'}} \, ,
\end{equation}
where the parameters $(\gamma', \kappa', z'_{p})$ parametrize the merger rate density as a function of redshift and are different than the (un-primed) parameters used to specify the star formation rate.  While the functional form given in Eq.~\eqref{eq:R_z_fit} provides a good fit to the merger rate density, the quantity $z'_{p}$ does not represent the peak of the merger rate.  To identify the maximum merger rate density, we can simply differentiate Eq.~\eqref{eq:R_z_fit} to obtain 
\begin{equation}
    z_\text{peak} = \bigg(\frac{\gamma'}{\kappa'-\gamma'}\bigg)^{1/\kappa'} (1+z'_p) - 1.
    \label{eq:zp_tilde}
\end{equation}
Finally, the probability distribution for redshift of BBH mergers can be written as,
\begin{equation}
    p(z_m) \propto \frac{1}{1+z_m} \left(\frac{dV_{\rm c}}{dz}\right)\mathcal{R}_m(z_m)
    \label{eq:p_z}
\end{equation}
where $V_{\rm c}$ represents the comoving volume and the factor $1/(1+z_m)$ accounts for redshifting the local merger rate to the observed merger rate.  While simulating the sources, the range for $z$ has been taken as [0, 10]. 

Figure~\ref{fig:input-pop-z} shows the redshift distribution of population of BBH mergers considered here, with a local merger rate of 22 Gpc$^{-3}$ yr$^{-1}$~\citep{KAGRA:2021duu}. The merger rate increases to redshift $\sim 1.5$ after which it falls rapidly.  The figure also shows the observable population of events in detector networks with sensitivity matching A+~\citep{O5c_sensitivity},  A\#, and next-generation observatories.  For the next-generation observatories, the vast majority of signals are observed, as shown in the right-hand plot in Figure~\ref{fig:input-pop-z}, which shows a detection efficiency greater than 60\% out to redshift of $z=10$.  For a network with A\# sensitivity the detection efficiency falls to 50\% by $z\approx 1$ and the peak of the observed redshift distribution is at $z\approx 0.7$ although signals are observed to redshifts of $z \approx 3$.  For the A+ sensitivity network, the observed redshift distribution peaks at $z \approx 0.4$ and the efficiency for drops below 10\% above $z=1$.  Consequently, an A+ network will have minimal sensitivity at the merger rate peak and hence we have not considered it in our subsequent analyses. 

In addition to the default Madau-Dickinson model $\psi_\text{MD}(z)$, with $\gamma = 2.7$, $\kappa=5.6$, and $z_p=1.9$, we consider two alternative star formation scenarios, with the peak of star formation moved to higher and lower redshifts.  We denote these \mdhigh, which has $z_p=2.53$, and \mdlow, with $z_p=1.54$. 
After the inclusion of time delay, the redshift distribution of these models peak at $z_\text{peak} \sim 1.2$ and $z_\text{peak} \sim 2$ for \mdlow\ and \mdhigh\ respectively. 
In Table~\ref{tab:num-sources}, we summarize the three population models and the expected number of observations in the GW detector networks under consideration. To calculate these detection rates, we impose a detection threshold on the optimal SNR of $\rho^\text{opt}>12$ for the network.  All of the parameter values for these models, as well as the fitted parameters $\gamma', \kappa', z'_p$, and $z_\text{peak}$ for the merger rate density are discussed in Appendix~\ref{app:fit_parameters}, with the parameters summarized in Table~\ref{tab:hyperpars}.  We also provide population and efficiency curves for \mdhigh\ and \mdlow\ models in Fig.~\ref{fig:input_pop_MDhigh_MDlow} of Appendix~\ref{app:fit_parameters}. The priors taken on population parameters for hierarchical Bayesian inference are mentioned in Table~\ref{tab:priors}. Previous works \citep[][for instance]{Vitale:2018yhm} have inferred the merger rate with additional parameters for time-delay but we take a simpler approach.

\subsection{Parameter estimation: Fisher matrix analysis}
\label{subsec:method:fm}

To infer the source parameters of a GW signal produced by a compact binary merger, Bayesian inference parameter estimation (PE) is usually performed, which requires sampling of the likelihood in a large-dimensional parameter space \citep{Ashton:2018jfp, pBilby, lalsuite, Romero-Shaw:2020owr, LIGOScientific:2025slb}.  However, this can be computationally expensive and time-consuming. A common workaround is to expand the log-likelihood in source parameters and truncate at quadratic order to provide a leading-order estimate of measurement uncertainties and correlations between parameters. The covariance of the resulting multidimensional Gaussian is given by the inverse of the Fisher information matrix \citep{Cutler:1994ys}. This approximation works well for high-SNR signals and breaks down at low SNRs.%
\footnote{The general expectation is that the Fisher Matrix (FM) approximation works well at high SNRs \citep[e.g.][]{Ajith:2009fz, Vallisneri:2007ev}. The exact value may depend on the information contained in the detected GW signal, the effective bandwidth (which is a measure of the number of cycles in band), and dimensionality of the problem. In general, the greater the dimensionality, the larger the SNR needs to be in order for the FM to produce meaningful constraints on the parameters \citep{Vallisneri:2007ev}.} 
In order to ensure the Fisher Matrix approximation works well for our analysis, we have chosen a detection threshold of a network SNR $\geq 12$ for our final events, ensuring we are not considering any very-low SNR signals. Moreover, although we allow the signals to have non-zero component spins, we do not vary the spins in our Fisher matrix analysis, as detailed in Step 2 below, so that the dimensionality of the Fisher matrix is kept at 9-dimensional instead of 15-dimensional. Finally, we reject the handful of points that result in poorly conditioned matrices.
In combination, these steps help ensure that we are employing the Fisher matrix approximation within its domain of validity.  We provide the details of the setup in the remainder of this section.

\subsubsection{Fisher Matrix formalism}

Let $d^{\mu}(t)$ be the detector strain in detector $\mu$, which consists of noise $n^{\mu}(t)$ and a CBC signal $h^{\mu}(t)$, where the observed signal depends upon the detectors' response to the gravitational-wave.  Then,
\begin{equation}\label{eq:data}
d^{\mu}(t) = n^{\mu}(t) + h^{\mu}(t).
\end{equation}
Assuming the noise to be stationary and Gaussian, the likelihood function for the binary's parameters $\boldsymbol{\theta}$ is given by: 
\begin{equation}
\mathcal{L}(\mathbf{d} \mid \boldsymbol{\theta}) \propto \prod_{\mu} e^{-\left(d^\mu - h^\mu(\boldsymbol{\theta})|d^\mu - h^\mu(\boldsymbol{\theta})\right)/2},
\label{eq:likelihood}
\end{equation}
where $(\cdot|\cdot)$ denotes the noise-weighted inner product \citep[see, e.g.][]{Cutler:1994ys}, and $\mathbf{d}$ represents data from the whole detector network. Expanding the log-likelihood to quadratic order about the peak of the distribution yields: 
\begin{equation}
\mathcal{L}(\mathbf{d} \mid\boldsymbol{\theta}) \propto e^{-\frac{1}{2}\Gamma_{ij}\Delta\theta^i \Delta\theta^j},
\label{eq:err_prob}
\end{equation}
where $\Delta \theta^i \equiv \theta^i - \bar{\theta}^i$, and $\bar{\theta}^i$ corresponds to the peak of the likelihood. Here, $\Gamma_{ij}$ is the Fisher information matrix for the network, which is defined as the sum over the individual Fisher matrices for each detector
\begin{equation}\label{eq:fisher_net}
    \Gamma_{ij} = \sum_\mu \Gamma^\mu_{ij}, 
\end{equation}
where
\begin{equation}\label{eq:fisher_det}
\Gamma_{ij}^{\mu} = \left(\frac{\partial h^\mu}{\partial \theta^i} \Big \rvert \frac{\partial h^\mu}{\partial \theta^j} \right).
\end{equation}
The Bayesian posterior probability for parameters $\boldsymbol{\theta}$, given the data $\mathbf{d}$ can be written as,
\begin{equation}
    p(\boldsymbol{\theta}|\mathbf{d}) = \frac{\mathcal{L}(\mathbf{d}\mid\boldsymbol{\theta})\pi(\boldsymbol{\theta})}{\mathcal{Z}(\mathbf{d})},
    \label{eq:bayes_theorem}
\end{equation}
where $\mathcal{L}(\mathbf{d}\ |\ \boldsymbol{\theta})$ is the likelihood given by Eq.~\eqref{eq:likelihood}, $\pi(\boldsymbol{\theta})$ is the prior probability distribution on the parameters $\boldsymbol{\theta}$, and $\mathcal{Z}(\mathbf{d})$ is the normalization of the posterior probability distribution, also known as the evidence,
\begin{equation}
    \mathcal{Z}(\boldsymbol{d}) = \int \mathrm{d} \boldsymbol{\theta} \mathcal{L}(\mathbf{d}|\boldsymbol{\theta})\pi(\boldsymbol{\theta}).
\end{equation}
Under the assumption of flat priors on the parameters $\boldsymbol{\theta}$, and using the Fisher approximation, the posterior distribution is
\begin{equation}
p(\boldsymbol{\theta} \mid \mathbf{d}) \propto  e^{-\frac{1}{2}\Gamma_{ij}\Delta\theta^i \Delta\theta^j} \, .
\end{equation}
The width of the posterior distribution is then given by the covariance matrix, $\Sigma$, which is defined as the inverse of the Fisher matrix,
\begin{equation}
\Sigma_{ij} = \langle \Delta \theta^i \Delta \theta^j \rangle = (\Gamma^{-1})_{ij}.
\end{equation}
In particular, the 1-$\sigma$ confidence interval of the posterior of parameters $\theta^i$, marginalised over all other parameters, is given by the square-root of the diagonal elements of the covariance matrix,
\begin{equation}
\sigma_i = \sqrt{\Sigma_{ii}} \, ,
\label{eq:one_sigma_err}
\end{equation}
and off-diagonal elements of the covariance matrix contain information about the correlation between different parameters.

\subsubsection{Identifying Observed events and redshift posteriors}

Having briefly introduced the Fisher Matrix formalism, we now describe the details of how it is used in this analysis. As has been discussed in \cite{Essick:2023upv}, unbiased population inference requires that the detection and parameter inference of individual events are dependent on the observed data (noise + signal) rather than the source parameters of the signal. The absence of noise in the signal leads to the detection and parameter inference being dependent on the source parameters of individual events, i.e. likelihood being centred at the source parameters. This breaks the underlying assumption in hierarchical inference of individual event likelihoods being independent as discussed in Sec. \ref{subsec:method:pop_inference}.

In order to satisfy this condition, we generate the Fisher matrix at the ``true'' parameters of the simulated signal, then use this to simulate the impact of stationary, gaussian noise --- both upon the detectability of the signal and the recovered parameters.  We then repeat the Fisher matrix calculation at these ``detected'' parameters in order to obtain posterior distributions for the parameters.  Throughout, we use the package \texttt{GWFish}~\citep{Dupletsa:2022scg} for Fisher matrix calculations. 

The steps used to identify events which would be observed in the GW detector network and obtain estimates of their parameters and associated uncertainties are as follows:

\begin{enumerate}
    \item We wish to identify those events which would be observed in the detector network.  We assume a threshold of 12 on the network matched-filter SNR, $\rho^{\mathrm{mf}} \geq 12$, obtained by filtering the noisy data. However, to reduce computational cost, we first calculate the expected, or optimal, network SNR, in the absence of noise, denoted $\rho^\text{opt}$.  We discard events with $\rho^\text{opt} < 10$ as these are unlikely to cross the detection threshold.%
\footnote{We have verified that events with $\rho^{\rm{opt}} < 10$ which cross the threshold of $\rho^{\rm{mf}} > 12$ after introduction of noise is less than 0.03\% for A\# and even lower for XG.}

\item We perform Fisher matrix analysis on those events with  $\rho^{\rm{opt}} \ge 10$.  For A\#, all events with $\rho^{\rm{opt}} \ge 10$ are used.  For XG networks, the number of events observed in one year of data is too large to use in the population inference routines and therefore we take a subset of randomly drawn events.  We choose this fraction so that we have similar number of events for the A\# and XG networks. 

The astrophysical population of events is generated with all 15 binary black hole parameters: component masses ($m_1, m_2$), luminosity distance ($d_L$), inclination angle ($\theta_\text{JN}$), time and phase of coalescence ($t_c, \phi_c$), sky location and polarization angles ($\alpha, \delta, \psi$), and spin parameters ($a_1, a_2, \theta_1, \theta_2, \phi_{12}, \phi_\text{JL}$).%
\footnote{Full details about parameter definitions can be found in Table E1 of \cite{Romero-Shaw:2020owr}.}
The Fisher matrix is constructed on a restricted 9-dimensional parameter space 
\footnote{The \texttt{GWFish}~\citep{Dupletsa:2022scg} package returns the Fisher matrix with respect to $\mathcal{M}_c$ and $d_L$.  However, particularly for high redshifts and high-mass systems, these parameters can be large, relative to the other parameters. Hence, even if the errors on these parameters are small, the Fisher matrix can become badly-conditioned due to a mismatch in the order of the values between different parameters, and hence cause high inversion errors when inverted to a covariance matrix. Therefore, after obtaining the Fisher matrix, we transform to $\ln \mathcal{M}_c$ and $\ln d_L$ to avoid this issue.}
\begin{equation}
    \bm{\theta} \equiv \{\ln \mathcal{M}_c, q, \ln d_L, \theta_\text{JN}, t_c, \phi_c, \alpha, 
    \delta, \psi \}.
\end{equation}
using the \texttt{GWFish} package~\citep{Dupletsa:2022scg}.
We use the waveform model \textsc{IMRPhenomXPHM}~\citep{Pratten:2020ceb} to generate waveforms in all steps. 

\item Using the Fisher matrix, we construct the 9-dimensional multivariate-Gaussian distribution for the event parameters, using the simulated values as mean values. We randomly select one sample per event, $\hat{\bm{\theta}}$, from this multivariate Gaussian distribution, to mimic the impact of stationary-gaussian noise in the detectors on parameter recovery.  This step removes the direct dependence of the inference on the true source parameters, which are unknown in a real GW event, and makes the simulated events realistic for population inference.

\item We calculate the event SNRs, $\hat{\rho}^\text{opt}$, using the new parameter values obtained above, $\hat{\bm{\theta}}$, and keep those events which meet the threshold $\hat{\rho}^\text{opt} \geq 12$. Since we have effectively introduced noise in the parameter values of each event in the previous step, this is similar to a detection threshold applied on matched-filter SNR $\rho^{\rm{mf}} \geq 12$.
    
\item Using this updated set of events, we perform Fisher matrix analysis again and construct a multivariate Gaussian distribution for the parameter distributions following step 2, but using the values $\bm{\hat{\theta}}$.%
\footnote{While a repeat Fisher matrix analysis is not strictly necessary if the new parameter values are close enough to the true values, some events included in our analysis have high redshift values which may result in change in error intervals and covariance matrices at the new parameter points.} 
Finally, we draw 10,000 redshift samples per event from these updated 9-dimensional multivariate Gaussian distributions to use as the inferred redshift distribution for the event.

\end{enumerate}

\subsection{Population inference}
\label{subsec:method:pop_inference}

Let $\mathcal{D} \equiv \{ d_1, d_2, ..., d_N \}$ be the data set for $N$ observed GW signals. Then the posterior probability distribution for the population parameters $\boldsymbol{\Lambda}$, marginalized over the merger rate $\mathcal{R}$ is given by \citep{Loredo:2004nn, Mandel:2018mve, Vitale:2020aaz}, 
\begin{equation}
    p( \boldsymbol{\Lambda}| \mathcal{D}) \propto \frac{1}{p^N_{\rm{det}}(\boldsymbol{\Lambda})}\frac{\mathcal{L}( \mathcal{D}|\boldsymbol{\Lambda}) \pi( \boldsymbol{\Lambda})}{\mathcal{Z}(\mathcal{D})},
    \label{eq:hyper_posterior}
\end{equation}
where $\mathcal{L}( \mathcal{D}|\boldsymbol{\Lambda})$ is the population likelihood, $\pi( \boldsymbol{\Lambda})$ denotes the prior on population parameters, and $\mathcal{Z}(\mathcal{D})$ is the evidence which is useful for comparing population models. In the above equation, $p^N_{\rm{det}}(\boldsymbol{\Lambda})$ encodes detector selection effects which we discuss in the next sub-section. Under the assumption that different observations are independent of each other, we can write the population likelihood as~\citep{Loredo:2004nn, Mandel:2018mve, Vitale:2020aaz}:\footnote{We have followed the notation for Eqs.~\eqref{eq:combined_likelihood}-\eqref{eq:p_det_theta} used in \cite{Singh:2023aqh}.}
\begin{equation}
    \mathcal{L}( \mathcal{D}|\boldsymbol{\Lambda}) = \prod_{i=1}^{N} \mathcal{L}(d_i| \boldsymbol{\Lambda}) = \prod_{i=1}^{N} \int \mathrm{d}\boldsymbol{\theta_i} \, \mathcal{L}(d_i| \boldsymbol{\theta_i}) \, \pi(\boldsymbol{\theta_i}|\boldsymbol{\Lambda}),
    \label{eq:combined_likelihood}
\end{equation}
where $d_i$ denotes the data for the $i^{th}$ event with parameters $\boldsymbol{\theta_i}$. Here $\pi(\boldsymbol{\theta_i}|\boldsymbol{\Lambda})$ denotes the population model prior for the distribution of $\boldsymbol{\theta}$ conditioned upon the population parameters $\boldsymbol{\Lambda}$. This assumes that all detected events have been drawn from the same population model. Substituting Eq.~\eqref{eq:combined_likelihood} in Eq.~\eqref{eq:hyper_posterior} results in the posterior on the population parameters $\boldsymbol{\Lambda}$ given by
\begin{equation}
    p( \boldsymbol{\Lambda}| \mathcal{D}) \propto \frac{\pi( \boldsymbol{\Lambda})}{p^N_{\rm{det}}(\boldsymbol{\Lambda})} \prod_{i=1}^{N} \int \mathrm{d}\boldsymbol{\theta_i} \ p(\boldsymbol{\theta_i}| d_i) \frac{\pi(\boldsymbol{\theta_i}|\boldsymbol{\Lambda})}{\pi(\boldsymbol{\theta_i})},
    \label{eq:hyper_pos_selec_effects}
\end{equation}
where we have used Bayes' theorem [Eq.~\eqref{eq:bayes_theorem}] to write the likelihood for individual BBH events in terms of the posterior, prior, and evidence. The multiplication of evidences $\prod_{i = 1}^{N} \mathcal{Z}(d_i)$ is absorbed in the proportionality as it is constant. The evaluations of integral in Eq.~\eqref{eq:hyper_pos_selec_effects} can be done by calculating the expectation value of the ratio of population prior, $\pi(\boldsymbol{\theta_i}|\boldsymbol{\Lambda})$, to the original parameter estimation prior, $\pi(\boldsymbol{\theta_i})$, for each event, and can be computed numerically via Monte-Carlo integration as:
\begin{equation}
    \int \mathrm{d}\boldsymbol{\theta_i} \ p(\boldsymbol{\theta_i}| d_i) \frac{\pi(\boldsymbol{\theta_i}|\boldsymbol{\Lambda})}{\pi(\boldsymbol{\theta_i})} \approx \frac{1}{N^i_{\rm{s}}} \sum_{j = 1}^{N^i_{\rm{s}}}  \frac{\pi(\boldsymbol{\theta_i^j}|\boldsymbol{\Lambda})}{\pi(\boldsymbol{\theta_i^j})}\Bigg\vert_{\boldsymbol{\theta_i^j}\sim p(\boldsymbol{\theta_i}| d_i)}
    \label{eq:monte_carlo_integration}
\end{equation}
where $\boldsymbol{\theta_i^j}$ denotes the $j^{th}$ sample drawn from the posterior of the $i^{th}$ event. The above approximation works well with sufficient number of posterior samples which, depending on the SNR of the events, can range between $\mathcal{O}(10^3 - 10^5)$.

\subsubsection{Selection effects}
\label{subsec:method:selection_effects}

The detection of GW signals depends primarily on the SNR of the signal.  This, in turn, depends on the parameters such as masses, spins, luminosity distance, inclination angle, and the sky location as well as the sensitivity of the detectors in the network.  This can lead to Malmquist bias \citep{1922MeLuF.100....1M, 1925MeLuF.106....1M} in the inference of the actual astrophysical population and is accounted for by a normalization factor, $p^N_\text{det}(\boldsymbol{\Lambda})$, in the population likelihood [Eq.~\eqref{eq:hyper_pos_selec_effects}]. The probability of detecting a binary drawn from an astrophysical population characterized by the parameters $\boldsymbol{\Lambda}$ can be computed by marginalizing the detection probability over the predicted population~\citep{Mandel:2018mve, Farr:2019rap, Vitale:2020aaz}.
\begin{equation}
    p_{\rm{det}}(\boldsymbol{\Lambda}) = \int \mathrm{d}\boldsymbol{\theta} p_{\rm{det}}(\boldsymbol{\theta}) \pi(\boldsymbol{\theta}| \boldsymbol{\Lambda}),
\end{equation}
where
\begin{equation}
    p_{\rm{det}}(\boldsymbol{\theta}) = \int_{d:\rho(d)\geq \rho_{\rm{th}}} \mathrm{d}d \ p(d | \boldsymbol{\theta}).
\label{eq:p_det_theta}
\end{equation}
In other words, the integration is carried out only on those likelihoods where data meets the detection criteria given by $\rho(d) > \rho_{\rm{th}}$ where $\rho_{\rm{th}}$ is the detection threshold. Computing $p_{\rm{det}}(\boldsymbol{\theta})$ numerically entails an extensive injection campaign where the detection statistic is calculated for simulated GW signals in the detector noise. Due a finite number of simulations, there is always an uncertainty in the selection function, although this can be mitigated by simulating enough GW signals \citep{Farr:2019rap, Essick:2022ojx, Talbot:2023pex, Essick:2023toz}.

We have used the same detection criterion for selection effects as the Fisher matrix calculation. $\mathcal{O}(10^7)$ injections were created and matched filter SNR for the network, $\rho^\text{mf}$, for each injection was simulated by drawing from a Normal distribution with $\mu=\rho^\text{opt}$ and $\sigma=1$. For population inference carried out in this work, we have used the \texttt{python} package \texttt{gwpopulation}~\citep{Talbot:2024yqw}.

\section{Results}
\label{sec:results}

\subsection{SNRs and event posteriors}
\label{subsec:res-snrs}

\begin{figure}
    \centering
    \includegraphics[width=\linewidth]{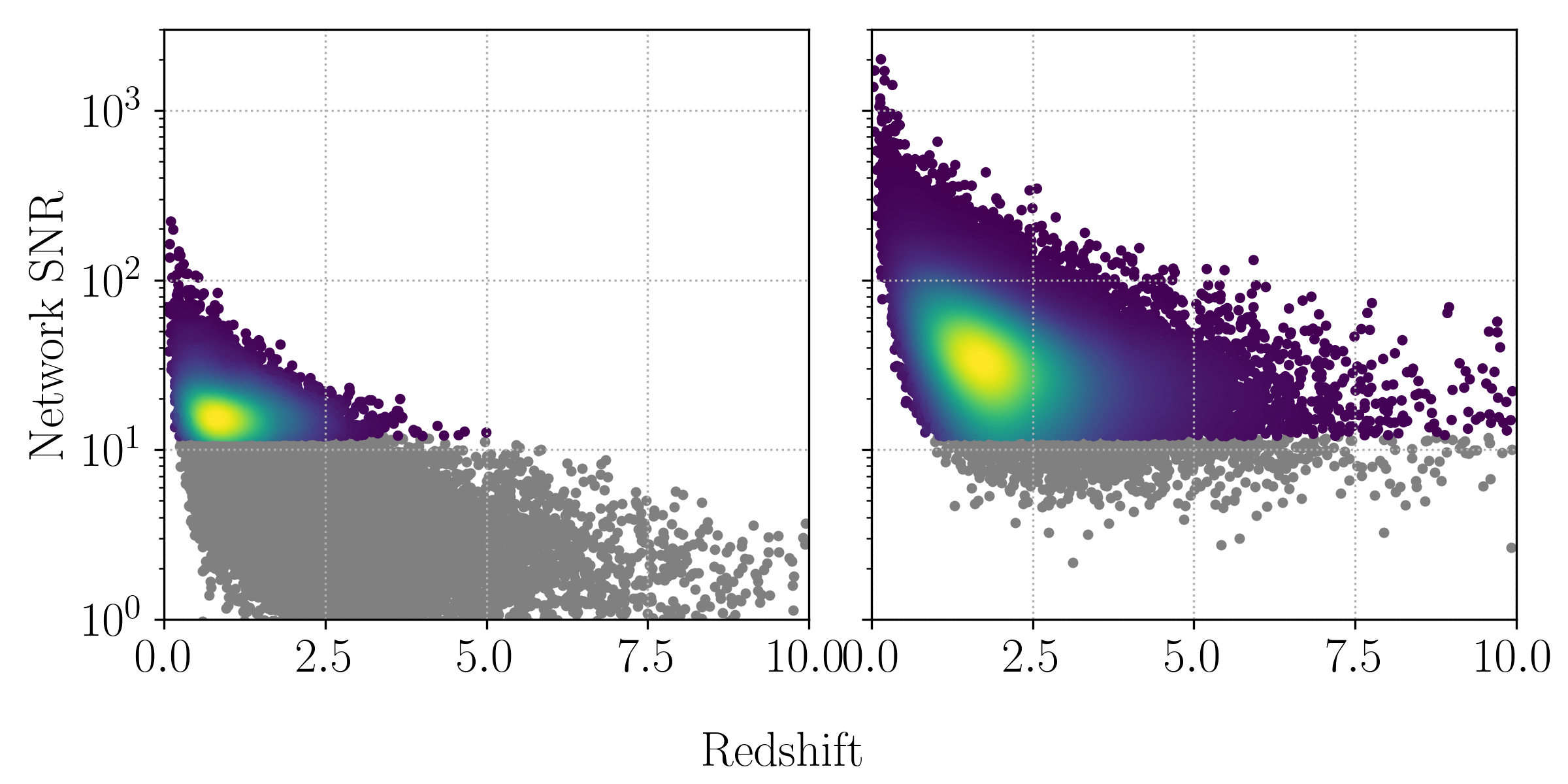}
    \caption{Network optimal SNR ($\rho^{\rm{opt}}$) plotted as a function of redshift for MD population in A\# (\textit{left}) and XG (\textit{right}) networks for 1 year of data. The events below the SNR of 12 have been greyed out, and the colour gradient for the detected events shows increasing point density from darker to lighter hues.}
    \label{fig:snr-vs-z}
\end{figure}

Figure \ref{fig:snr-vs-z} shows the SNR distribution of detected events for the MD population as a function of redshift, where the dark to light hues indicate the density of points going from lowest to highest. As expected from Figure \ref{fig:input-pop-z}, the majority of signals are observable in the XG network, and the peak density occurs at $\rho \sim 30$ at a redshift of $z \sim 1.5$.  Therefore, a typical signal at or close to redshift of the peak merger rate will be well above the detection threshold and the parameters will likely be recovered with good accuracy.  In contrast, for the A\# sensitivity network, the peak of the observed distribution occurs closer to $z=1$ and with SNRs close to the detection threshold.  Furthermore, there are a large number of events which are undetectable.

\begin{figure}
    \centering
    \includegraphics[width=\linewidth]{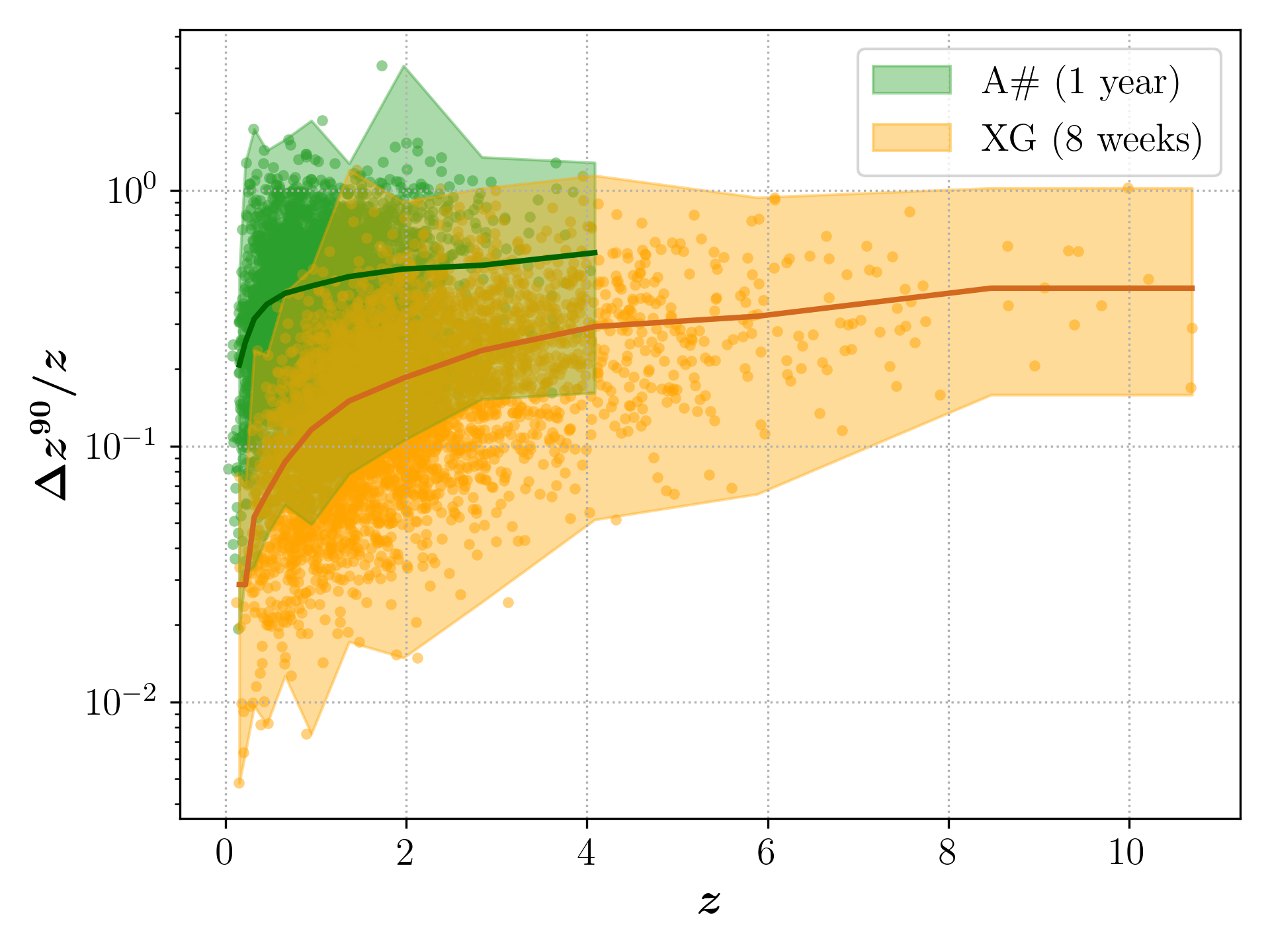}
    \caption{90\% uncertainty intervals for redshift posteriors relative to median values of the posteriors, plotted as a function of the median values, for A\# and XG networks, for MD population. The shaded regions represent the minimum and maximum regions of the respective redshift error bin, and the solid lines denote the mean values for the errors in each bin. The time interval mentioned in the labels indicates the equivalent time of observation for each detector network corresponding to the number events analysed here. Although we have analysed $\sim 8$ weeks of data for XG, the Fisher errors for 1 year of data will be similar since the subset has been randomly chosen from a year's worth of population.}
    \label{fig:deltaz-vs-z}
\end{figure}

\begin{figure*}
    \centering
    \includegraphics[width=0.32\linewidth]{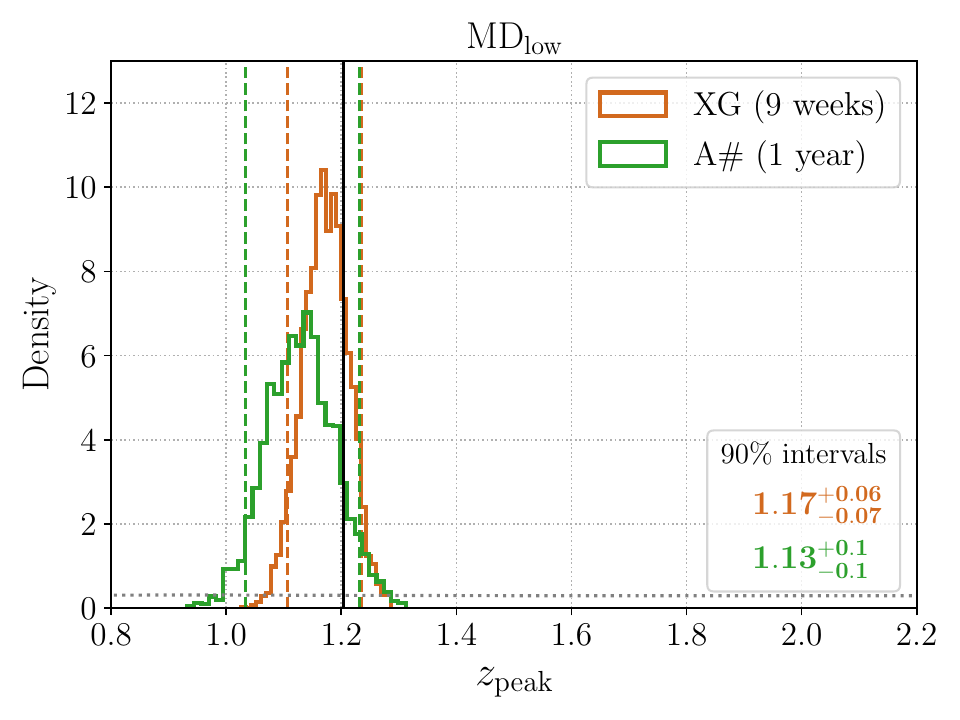} \includegraphics[width=0.32\linewidth]{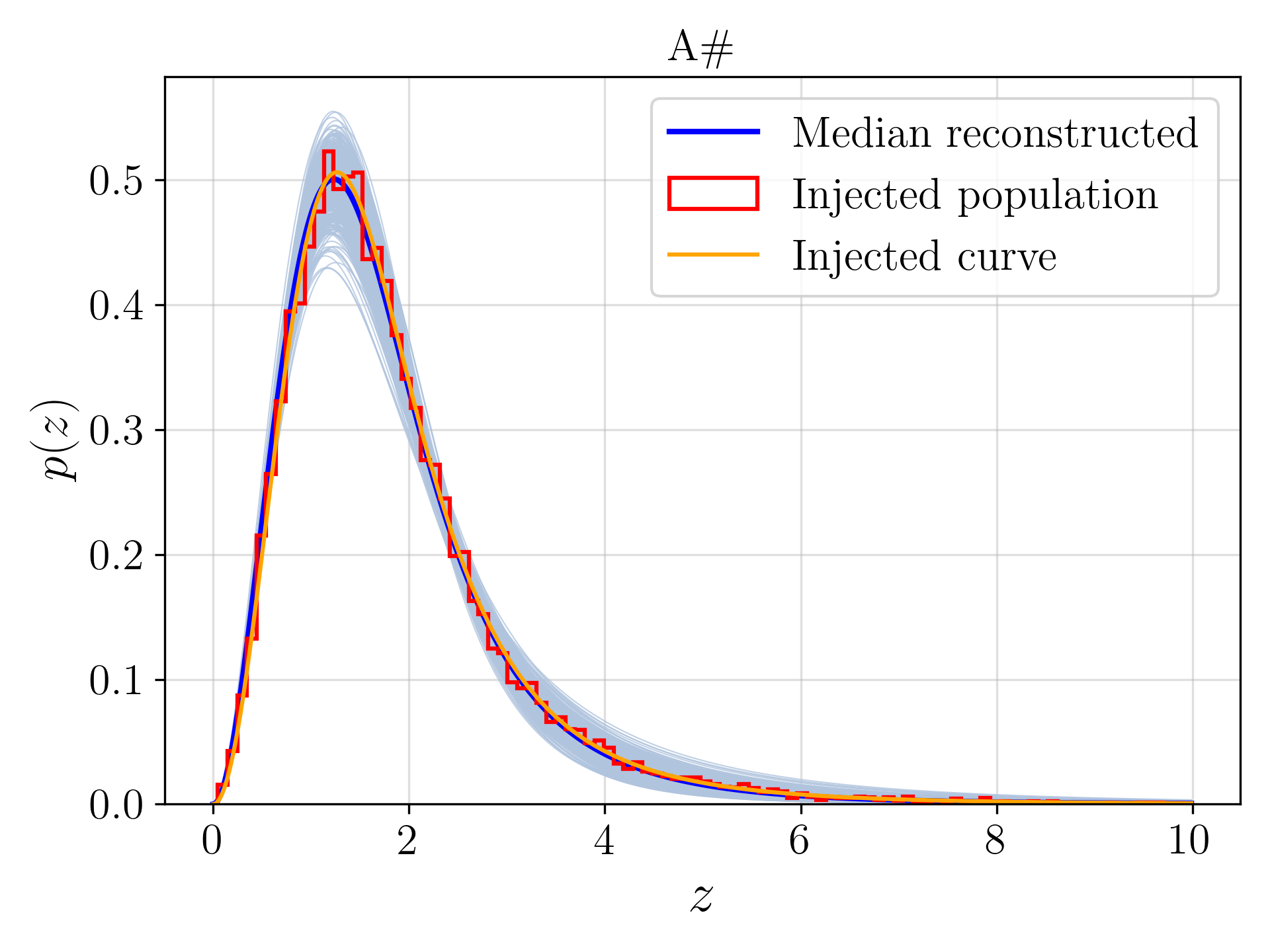}    \includegraphics[width=0.32\linewidth]{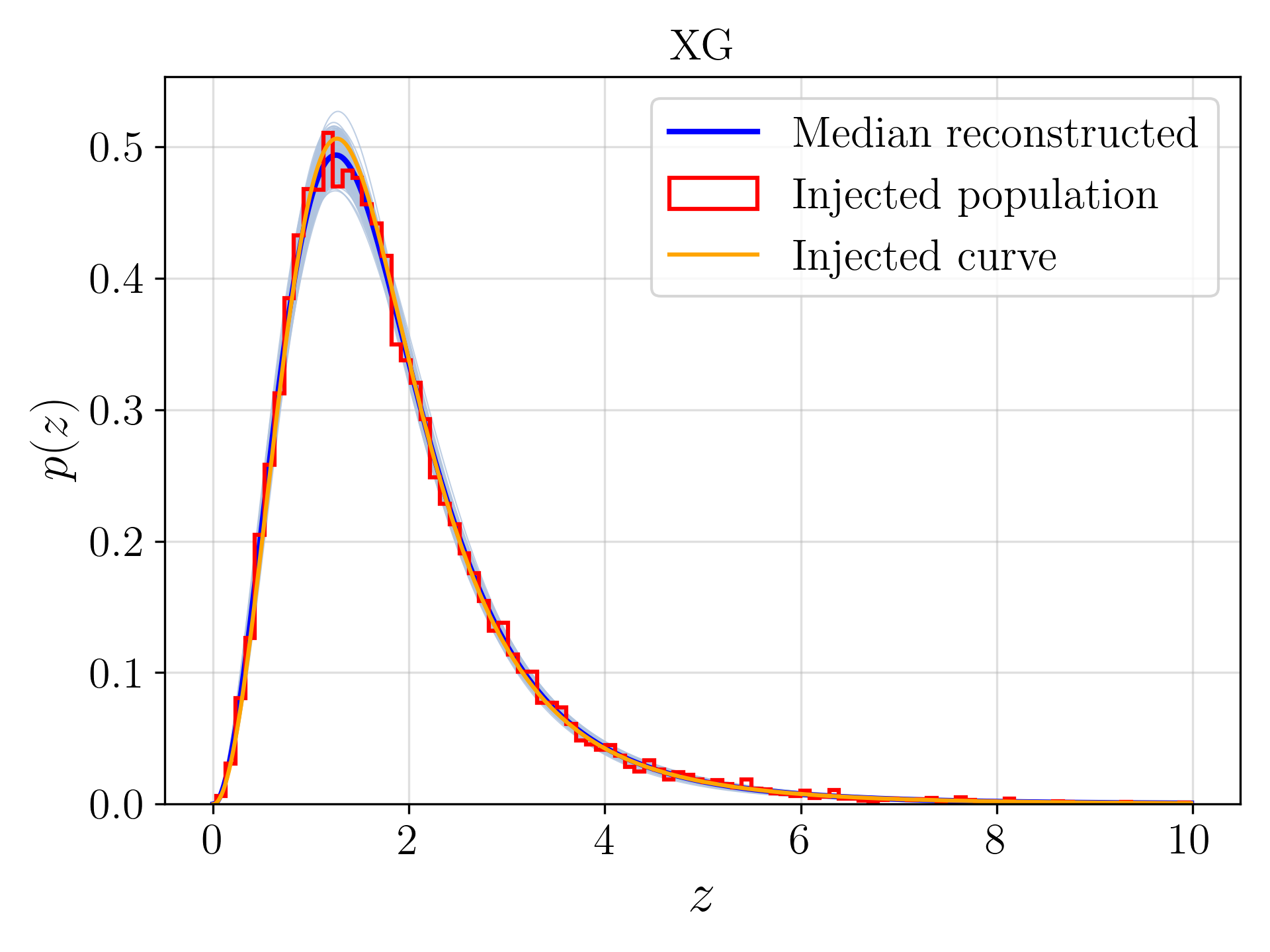}
    
    \includegraphics[width=0.32\linewidth]{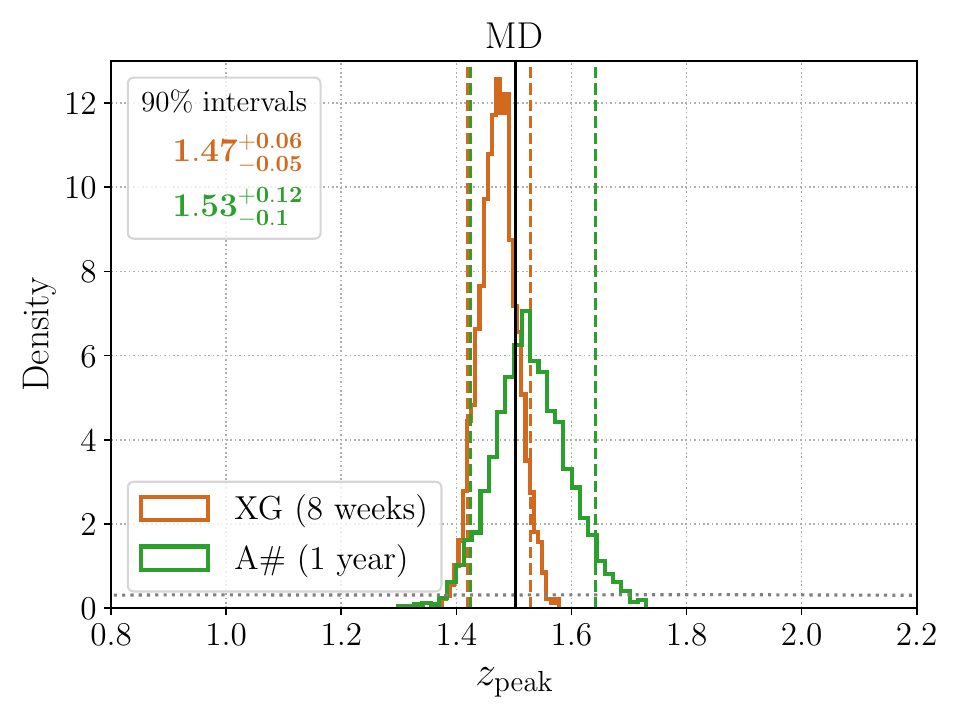}
    \includegraphics[width=0.32\linewidth]{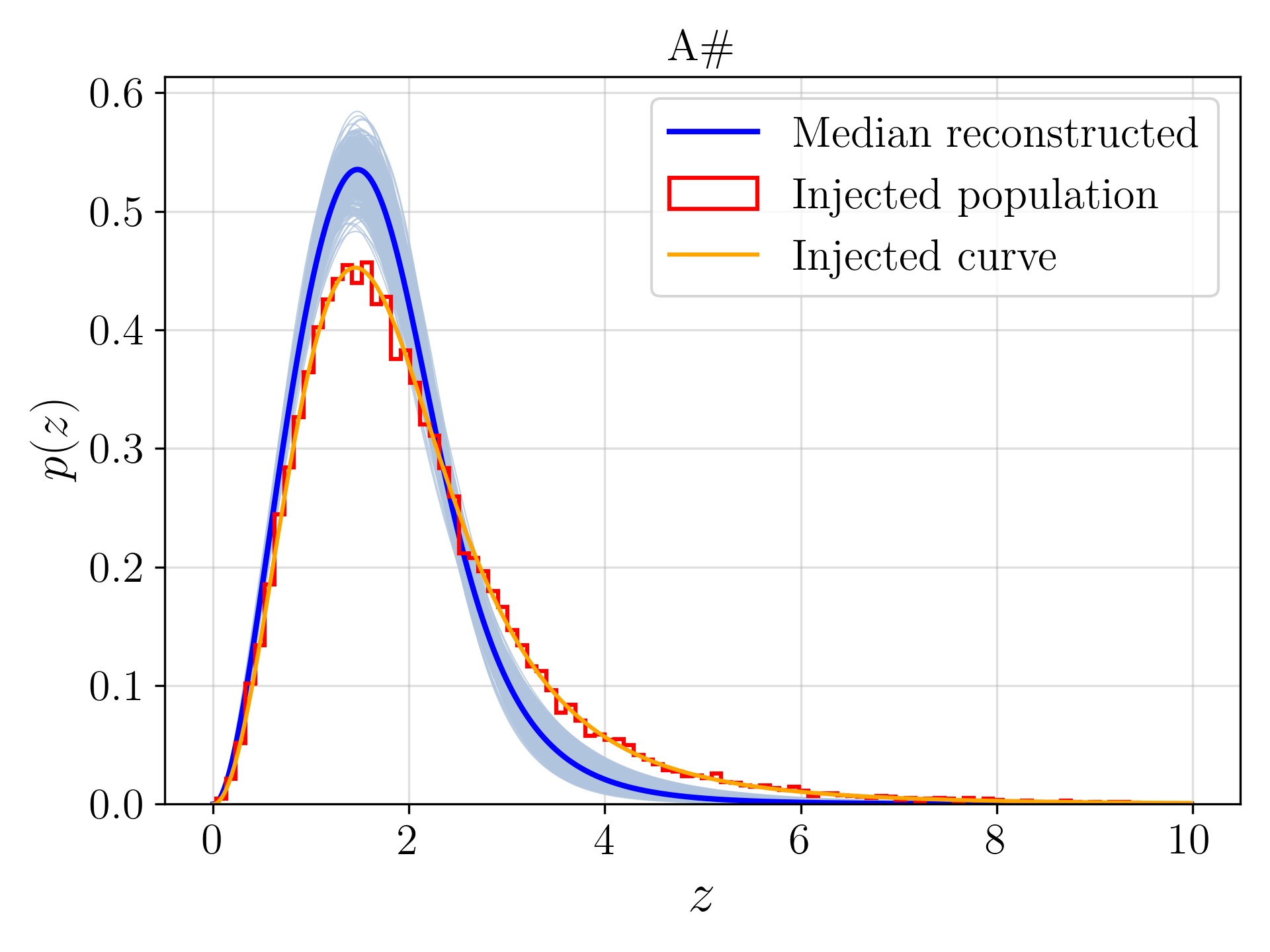}    \includegraphics[width=0.32\linewidth]{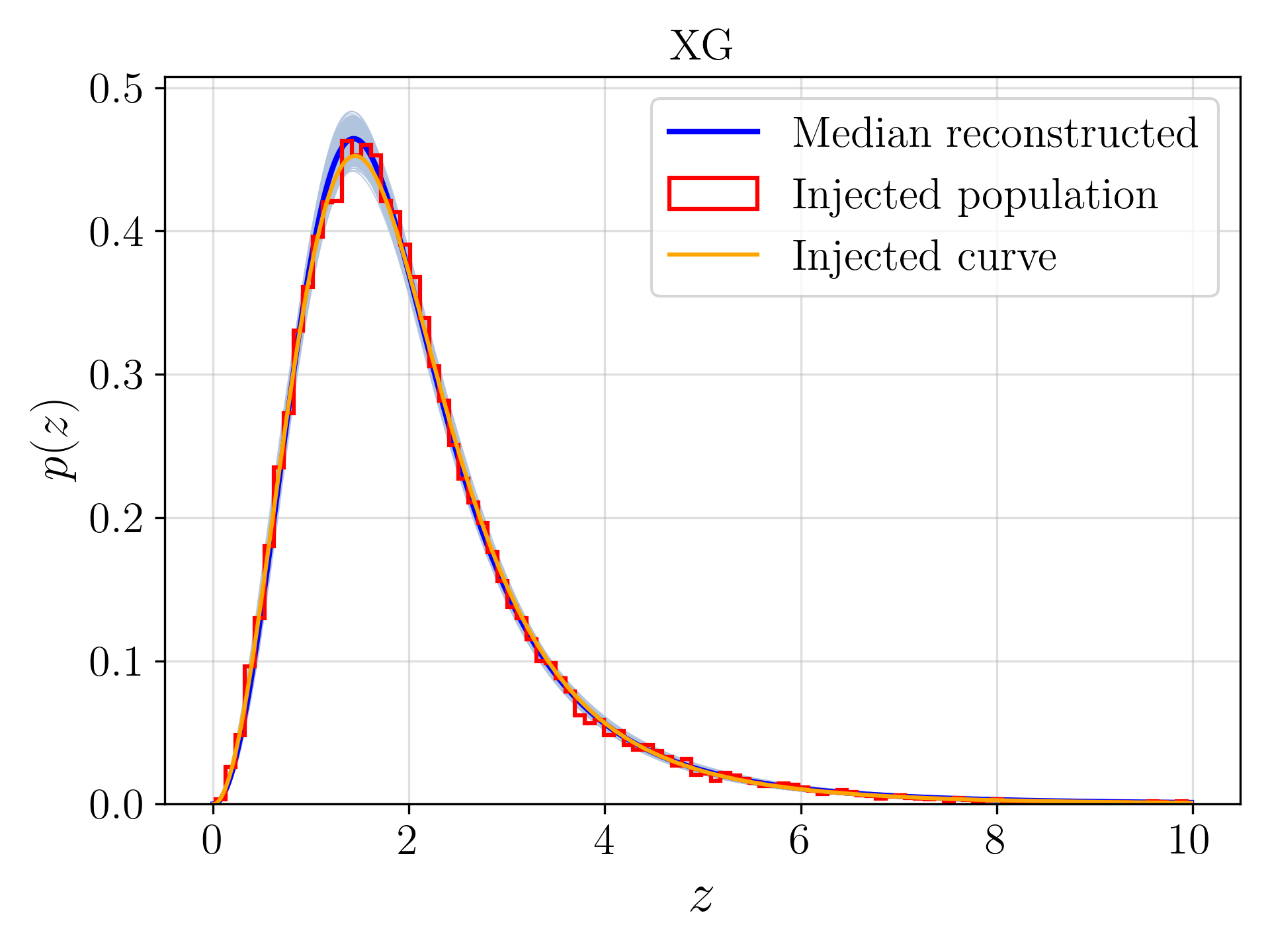}
    
    \includegraphics[width=0.32\linewidth]{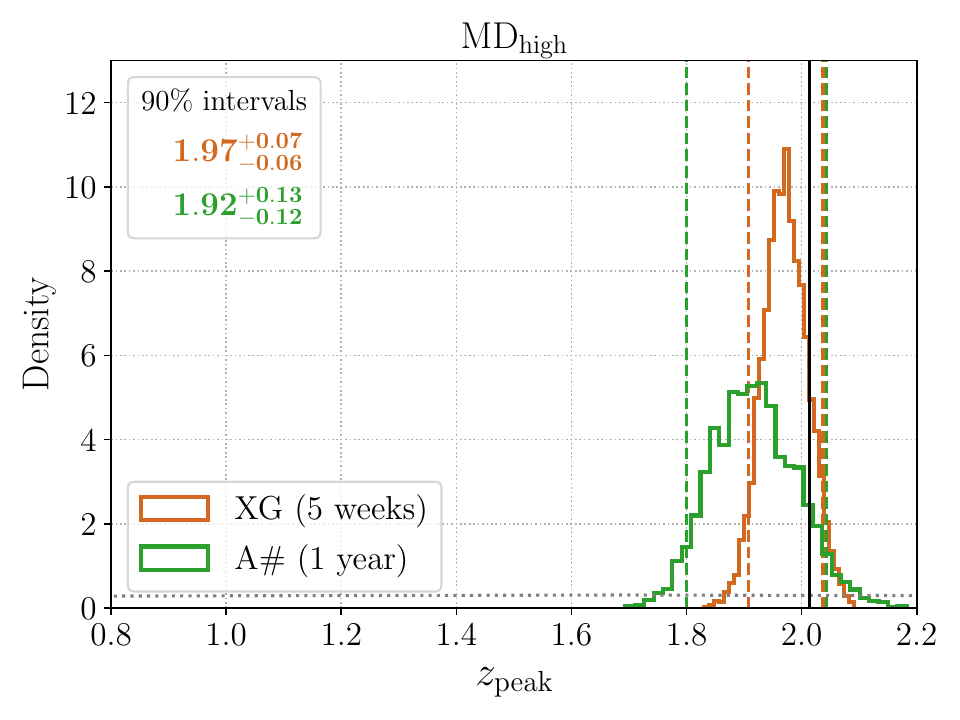}    \includegraphics[width=0.32\linewidth]{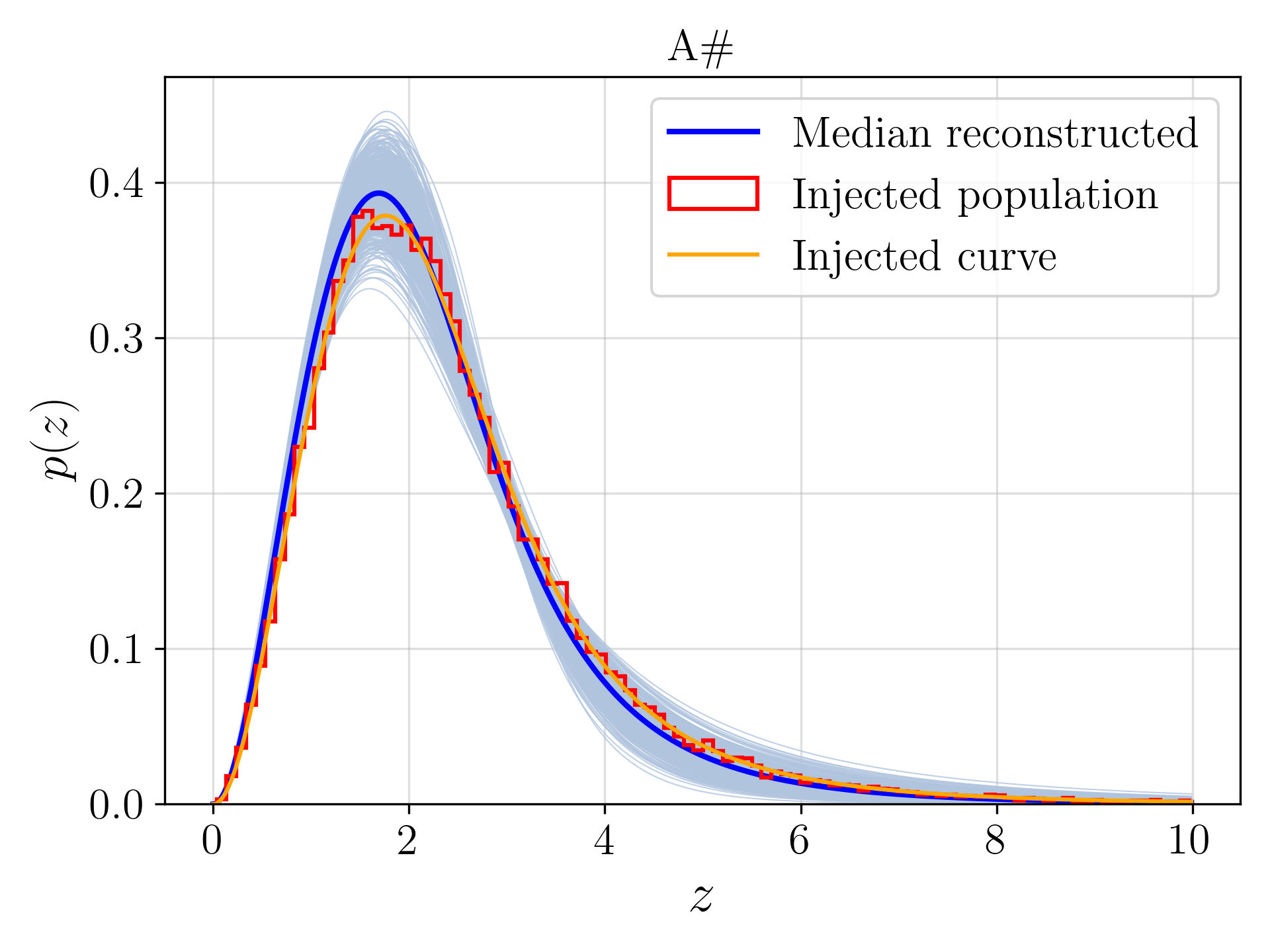} \includegraphics[width=0.32\linewidth]{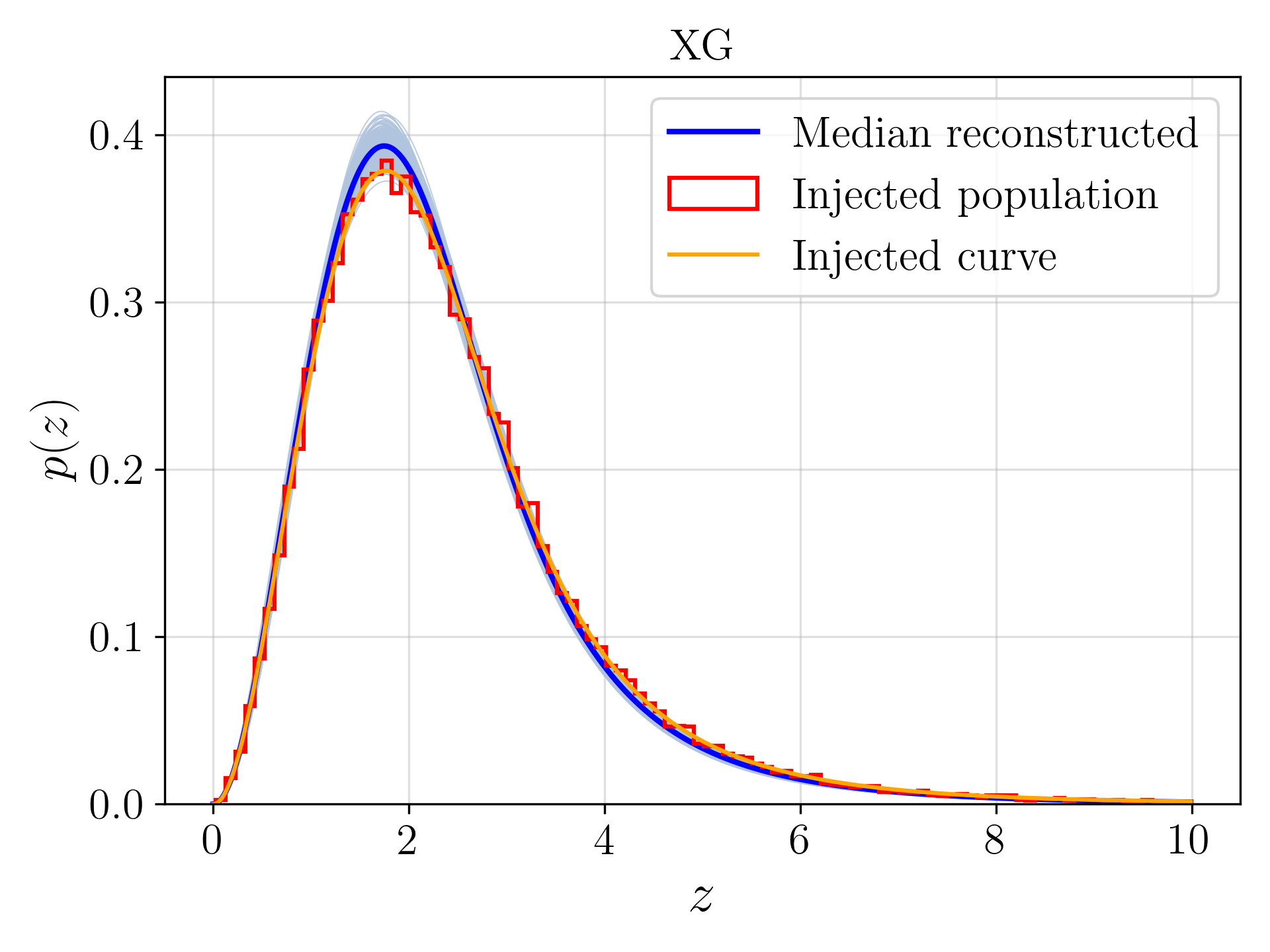}
    \caption{Left: Posteriors on $z_\text{peak}$ for \mdlow\ (top), MD (middle), and \mdhigh\ (bottom) populations for the XG (green) and A\# (orange) networks. The middle and right columns show the reconstructed redshift distribution $p(z)$ [Eq.~\eqref{eq:p_z}] after population inference for A\# and XG networks, respectively. Dark blue curves denote the reconstruction of $p(z)$ when median values of samples ($\gamma', \kappa', z_p'$) are taken. The thin blue lines correspond to 500 random samples taken from these posteriors to show the error bars. For reference, we have also plotted the histogram of the input population (as shown in Fig.~\ref{fig:input-pop-z}) in red, along with the functional form of $p(z)$ in light orange in each case.}
    \label{fig:hist_total}
\end{figure*}

\begin{figure}
    \centering
    \includegraphics[width=\linewidth]{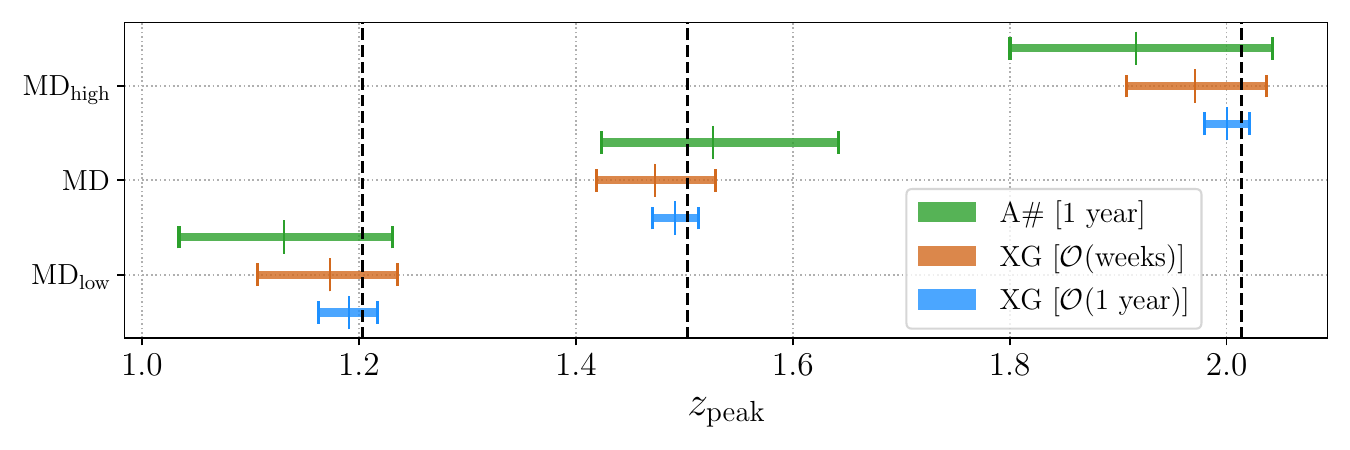}
    \caption{90\% confidence intervals for $z_\text{peak}$ posteriors for A\# (shown in green) and XG (shown in brown and blue). The error interval plotted in brown corresponds to the 90\% interval obtained when same number of events are analylsed in XG network as A\# which corresponds to 9, 8, and 5 weeks of equivalent observation time in XG for \mdlow, MD, and \mdhigh\ models respectively. The error interval in blue, on the other hand, is an indicative interval drawn by scaling the brown curve for 1 year of detections in XG.}
    \label{fig:ci_scaled_total}
\end{figure}

Following the Fisher matrix analysis, we construct posteriors for redshift following the steps described in Sec.~\ref{subsec:method:fm}. Figure \ref{fig:deltaz-vs-z} shows the 90\% uncertainty regions for the inferred redshift plotted against the median values.  The solid lines denote the mean values for the uncertainty in the measured redshift while the shaded regions show the minimum and maximum redshift uncertainty in different redshift ranges.  We show the results for both the A\# and XG networks. As expected, the XG network is sensitive to signals at greater redshifts, so the observed population extends to the edge of our simulation at $z=10$ while for A\#, the most distant observed signal is at $z=4$.  Further, the redshift uncertainty is smaller at low redshifts, and increases at higher redshift as the signals will be observed with lower SNR. At high redshift, the uncertainty in the measurement of the redshift of a single system is typically tens of percent, and can be as large as 100\%. At a fixed redshift, the XG network measures the redshift with significantly greater precision, ranging from a factor of two to almost a factor of 10.

\subsection{Population inference}
\label{subsec:res:pop-inf}

\begin{figure}
    \centering
    \includegraphics[width=\linewidth]{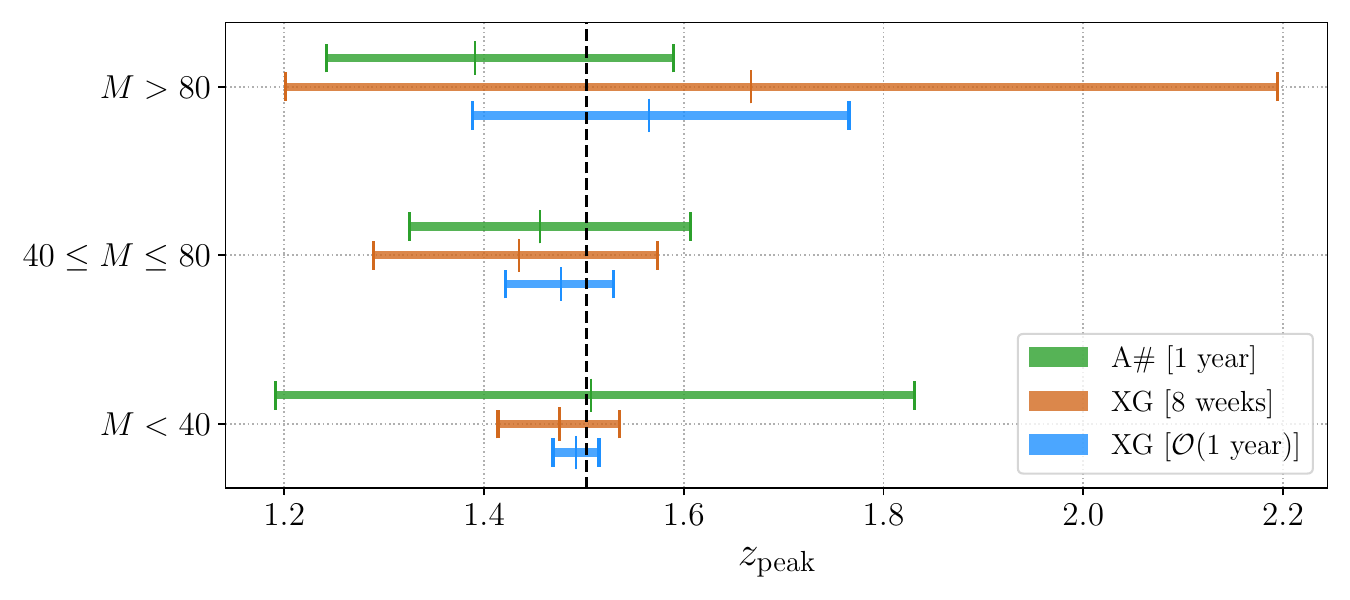}
    \caption{90\% confidence intervals for $z_\text{peak}$ posteriors for A\# (shown in green) and XG (shown in brown and blue) for various mass bands in MD population. The error intervals plotted in brown corresponds to the 90\% interval obtained when an equivalent of eight weeks of data is analysed for XG. The error interval in blue, on the other hand, is an indicative interval drawn by scaling the brown curve for one year of detections in XG.}
    \label{fig:ci_scaled_mass_bands}
\end{figure}

\begin{figure}
    \centering
    \includegraphics[width=\linewidth]{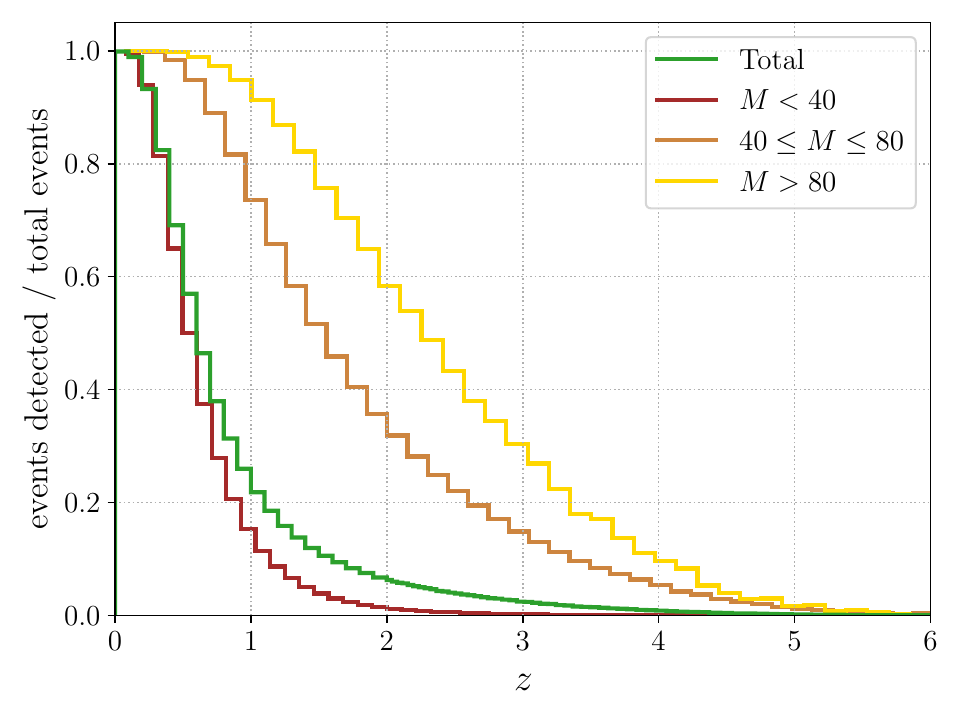}
    \caption{Efficiency curve for A\# network for various mass bands.}
    \label{fig:eff_mass_bands}
\end{figure}

Fig.~\ref{fig:hist_total} summarizes the results from our population inference study for the two detector networks, A\# and XG and the three different population models. The left column of Fig.~\ref{fig:hist_total} shows the inferred distribution for the redshift of peak merger rate, $z_\text{peak}$, for \mdlow\ (top), MD (middle), and \mdhigh\ (bottom) populations comparing the XG and A\# networks. As mentioned in Step (i) in Sec.~\ref{subsec:method:fm}, the number of events detected in XG is several times the number detected in A\# network. Performing population inference on all detected events is computationally expensive, hence we have chosen a number for XG that is similar to a year's worth of detections in A\#. This number is equivalent to a few weeks of observations for XG (exact numbers are slightly different for different population models and given in Table~\ref{tab:num-sources}). Moreover, if a year's worth of detections is considered both for XG and A\#, it is obvious that the posteriors on population parameters obtained in XG would be narrower than A\# due to sheer number of events. By choosing comparable number of events for both networks, we are able to explore if the XG network performs better than A\# solely due to the quality of events (higher SNRs in XG compared to A\#). It can be seen in Fig.~\ref{fig:hist_total} that even with the same number of events in both the detector networks, XG performs better than A\#. For example, across the three simulated populations, the XG network is able to recover the redshift with approximately half the uncertainty of the A\# network, \textit{for a comparable number of events}.  This is expected, based on the improved accuracy of measurement for individual events, as shown in Fig.~\ref{fig:deltaz-vs-z}. 

The middle and right columns of Fig.~\ref{fig:hist_total} show the reconstructed redshift distribution $p(z)$ [Eq.~\eqref{eq:p_z}] for the A\# and XG networks, respectively. Dark blue curves denote the reconstruction of $p(z)$ when median values of samples ($\gamma', \kappa', z_p'$) are taken. The thin blue lines correspond to 500 random samples taken from these posteriors to show the error bars. For reference, we have also plotted the histogram of the input population (as shown in Fig.~\ref{fig:input-pop-z}) in red, along with the functional form of $p(z)$ in light orange in each case. In addition to a tighter bound on the $z_\text{peak}$ parameter, the overall redshift distribution is also much better constrained for XG. It can be noted that while the position of the peak is correctly recovered in A\#, the other parameters, especially $\kappa'$ which determines the fall of the curve at redshift above $z_{\mathrm{peak}}$, are poorly constrained due to very few sources beyond the peak. This contributes to a poor recovery of the overall shape of the reconstructed curve for A\# which is not the case for XG.

While Fig.~\ref{fig:hist_total} compares XG and A\# for the same number of events, in Fig.~\ref{fig:ci_scaled_total} we scale the 90\% error intervals for XG to provide illustrative results for a year's worth of data. Fig.~\ref{fig:ci_scaled_total} shows the 90\% error intervals for A\# in green, error intervals for equal number of events in XG in brown, and an indicative 90\% error interval for 1 year of XG data in blue. We have obtained blue error bars by multiplying the error intervals in orange with $\sqrt{N_\text{analysed}/N_\text{total}}$ where N is the number of events. In addition to scaling the error interval, we have also scaled the offset between the median and the true value by the same factor, hence shifting the median accordingly.

\subsection{Specific mass ranges}
\label{subsec:mass-bands}

\begin{table*}
\begin{tabular}{|c|c|cc|cc|}
\hline
\multirow{2}{*}{\textbf{Mass Bin}} &
  \multirow{2}{*}{\textbf{\begin{tabular}[c]{@{}c@{}}Total sources in mass bin\\ (\% of the \\ total population)\end{tabular}}} &
  \multicolumn{2}{c|}{\textbf{Detection percentage}} &
  \multicolumn{2}{c|}{\textbf{\begin{tabular}[c]{@{}c@{}}No. of sources analysed\\ (equivalent observation time)\end{tabular}}} \\ \cline{3-6} 
 &
   &
  \multicolumn{1}{c|}{\textbf{A\#}} &
  \textbf{XG} &
  \multicolumn{1}{c|}{\textbf{\begin{tabular}[c]{@{}c@{}}A\#\\ (1 year)\end{tabular}}} &
  \textbf{\begin{tabular}[c]{@{}c@{}}XG\\ ($\sim$8 weeks)\end{tabular}} \\ \hline
\mlow  & \begin{tabular}[c]{@{}c@{}}26938\\ (85\%)\end{tabular} & \multicolumn{1}{c|}{7.5\%}  & 95.5\% & \multicolumn{1}{c|}{2015} & 3718 \\ \hline
\mmid  & \begin{tabular}[c]{@{}c@{}}4494\\ (14\%)\end{tabular}  & \multicolumn{1}{c|}{41\%}   & 99.8\% & \multicolumn{1}{c|}{1839} & 647  \\ \hline
\mhigh & \begin{tabular}[c]{@{}c@{}}296\\ (1\%)\end{tabular}    & \multicolumn{1}{c|}{86.5\%} & 100\%  & \multicolumn{1}{c|}{256}  & 35   \\ \hline
\end{tabular}
\caption{Distribution of sources across three mass bins for MD population model. Columns 3\textsuperscript{rd} and 4\textsuperscript{th} indicate the detectability of the respective networks for particular mass bins and the last two columns indicate the actual number of events analysed while performing population inference for each of the detector networks.}
\label{tab:num-sources-mass-bins}
\end{table*}

Next, we wish to explore the contribution of BBH mergers from different mass ranges in constraining the merger peak. While the low mass BBH mergers dominate the population in number, the high mass mergers are higher in SNR hence produce tighter posteriors on event redshifts. Moreover, due to a difference in metallicity, time-delay distributions, and formation channels, binaries in different mass bands may result in different peaks. While we assume the position of the peak to be the same across different mass bins in this work, our aim is to explore how well that peak can be constrained. Hence, we further divide the event posteriors analysed above into three mass bands: \mlow, \mmid, and \mhigh\ where $M$ is the total mass of the binary in source frame. Whether an event falls in one of these mass categories is determined by whether the 50th percentile for the source-frame total mass ($M$) posterior of that event falls in the respective category.

Fig.~\ref{fig:ci_scaled_mass_bands} shows the 90\% confidence intervals for $z_\text{peak}$ for the A\# and XG networks. Essentially, we have taken the events from the MD model, analysed and shown in Fig.~\ref{fig:ci_scaled_total}, and divided them into various mass bins and performed the population inference separately for these sub-categories. This means that for A\#, one year of data was analysed for the respective mass bins, whereas for XG, again, an equivalent of eight weeks of data were analysed. Now, upon closer inspection, this translates to a very different number of events across the two networks. First, the mass model (\textsc{Power Law + Peak}), by design, ensures that $\sim 85\%$ of events of the total population fall in the lowest mass bin (\mlow). The other two mass bins (\mmid) and (\mhigh) get $\sim 14\%$ and $\sim 1\%$ of events respectively. Second, A\# has poor detectability ($7.5\%$) in the low mass bin, better ($41\%$) in the mid mass bin, and more than 85\% in the high mass bin, whereas XG detects over 95\% of events in all mass bins.  This means that when one year of data is analysed for A\#, the number of events analysed is comparable between the low and mid mass bins while for XG the vast majority of events are in the low mass bin. We have summarized these numbers in Table~\ref{tab:num-sources-mass-bins}.   

Figure~\ref{fig:ci_scaled_mass_bands} shows the 90\% error intervals for $z_\text{peak}$. The green error bars show confidence intervals for A\#. Even though there are comparable numbers of events in both low and mid-range mass bands, the error band for \mlow\ is approximately twice as wide as \mmid. This is because the events in the (\mmid) band are higher in mass and hence have higher SNR, and thus have tighter posteriors on the redshift parameter. Further, the events in the (\mhigh) band are even fewer in number but again have high SNRs, which translates to good posteriors on redshift, hence comparable bounds on $z_\text{peak}$ for A\#. This can also be understood by looking at the efficiency curve for A\# network for various mass bands as shown in Fig.~\ref{fig:eff_mass_bands}. 

For XG, however, these numbers are widely different. Since XG has nearly perfect ($>95\%$) detectability in all mass bands, the number distribution of events is the same as in the input population. These numbers are reflected well in the orange and blue bands shown in Fig.~\ref{fig:ci_scaled_mass_bands}. Orange bands represent the confidence intervals for the equivalent of eight weeks of observation time of the XG network, whereas blue bands are indicative of $\mathcal{O}(1 \text{year})$ of observation time and have been obtained by scaling the orange bands by the appropriate factor. For XG, the observed events span larger redshift ranges and larger SNR range as shown in Fig.~\ref{fig:snr-vs-z}, and the redshift errors have a wide range of values as shown in Fig.~\ref{fig:deltaz-vs-z}. Therefore, the confidence intervals obtained for XG networks in blue are primarily determined by the number of events. Consequently, we have the smallest confidence interval for \mlow, which has the largest number of events, followed by \mmid\ and \mhigh. It is also worth noting that 8 weeks of data for \mhigh\ translates into very few events, just 35, while the other mass bands have hundreds or thousands of events, hence the error bar for \mhigh\ is large compared to the other error bars shown in Fig.~\ref{fig:ci_scaled_mass_bands}. Finally, although the inferred uncertainty for \mhigh\ with one year of XG data is comparable to the A\# result, this result is likely impacted by scaling the small number of events in the XG simulation. In general we would expect that XG observations would result in better $z_\text{peak}$ constraints than A\# due to the higher efficiency and greater SNR observed in XG detectors.  Nonetheless, it is clear that for \mhigh, it is the small number of events, and not the detector sensitivity, which has the strongest impact on our ability to infer $z_\text{peak}$.


\section{Discussion} 
\label{sec:concl}

In this work we analysed a population of BBH sources from three different population models for redshift distribution. We considered the standard Madau-Dickinson redshift distribution with an inverse time-delay function along with two additional fiducial Madau-Dickinson models where the peak of the distribution was placed at lower and higher values than the standard ($z_p=1.9$). We performed population inference on a year's worth of detections for the A\# network and a similar number for XG network to obtain bounds on the population parameters for redshift distribution. We observed that even with the A\# detector network, the peak of the redshift distribution is constrained well. With the XG detector network, the bounds on population parameters are much tighter, even when the same number of events are analysed as A\#, and the overall shape of the redshift distribution is also well recovered with much smaller error bands.

With the current LIGO-Virgo-KAGRA detectors, it is nearly impossible to constrain the star formation rate and distribution due to the limited reach of these detectors. It is generally expected that with the next-generation detectors such as Cosmic Explorer and Einstein Telescope, this should be easily doable. But these detector networks may still be at least a decade away from operation. In this work, we highlight the importance of the LIGO-A\# network and demonstrate that a network of A\# detectors is able to constrain the peak of BBH mergers with $\mathcal{O}(0.1)$ accuracy. Moreover, since the reach of A\# is up to $z \approx 4$, recovering the position of the peak (if it lies) within that range should aid in refining some of the underlying assumptions that go into SFR calculations from EM observations. 

For instance, if the merger peak is observed to be close to or higher than the SFR peak reported by the EM observations, it can point to the need for additional data points in galaxy catalogs at higher redshifts and a revisit of the selection effects taken into account. It could also mean that high mass BHs are preferentially produced in low metallicity galaxies which would shift the BBH peak to a higher redshift than the SFR peak. On the other hand, a peak which is significantly lower than the SFR peak would demand an update in the time-delay distributions. 

Furthermore, these same astrophysical effects might lead to a merger rate peak that varies across the BBH mass spectrum.  In this case, it is interesting to investigate whether a mass dependence of $z_{\mathrm{peak}}$ would be obervable with the A\# network. As a straightforward illustration of this, we split the events into three mass bands depending on the source total mass and observe that for A\# the peak is accurately constrained for higher mass systems $M > 40 M_{\odot}$.  For lower masses, the sensitivity of A\#, means that only a small fraction of events around $z_{\mathrm{peak}}$ are observable leading to reduced measurement accuracy.

It should, however, be noted that the data for GW observations of BBH mergers may include a number of formation channels which may result in an abundance of BBH binaries at certain redshifts rather than a straight-forward evolution from the SFR curve. With XG network, this should be easier to resolve since BBH mergers at higher redshifts will also be accessible and we will be able to constrain and comment on the overall shape of the merger rate curve. For instance, a slower fall of the curve at higher redshift (than expected from SFR curve) would indicate that our assumptions about the initial mass function (IMF) may need modification. Moreover, with XG, the constraints on other BBH parameters, such as masses, spins and eccentricity, are also expected to improve, in addition to estimates on luminosity distance. These parameters can play a critical role in distinguishing the various formation channels at different redshifts and will enable us to, e.g., study the mass dependence of the evolution of merger rate with redshift. Overall, GW observations in the next-generation detectors will serve as an independent source of observations complementing the galaxy surveys which have placed various constraints on the star-formation history in the past few decades.

\section*{acknowledgments}

We thank Aditya Vijaykumar for useful comments on the manuscript. We thank Paul Clark for useful discussions. DJ, SF, MH, and MKS acknowledge the Science and Technology Facilities Council (STFC) for support through grants ST/V005618/1 and ST/Y004272/1. Computations were performed on the CIT cluster provided by the LIGO Laboratory. The authors are grateful for computational resources provided by the LIGO Laboratory and supported by National Science Foundation Grants Np.~PHY-0757058 and No.~PHY-0823459. We used the following software packages: {\tt LALSuite}~\citep{lalsuite}, {\tt PyCBC}~\citep{alex_nitz_2020_4134752}, {\tt NumPy}~\citep{Harris:2020xlr}, {\tt Matplotlib}~\citep{2007CSE.....9...90H}, {\tt Seaborn}~\citep{Waskom2021}, {\tt jupyter}~\citep{soton403913}, {\tt dynesty}~\citep{Speagle:2019ivv}, {\tt corner}~\citep{corner}, \texttt{bilby}~\citep{Ashton:2018jfp}, \texttt{gwpopulation}~\citep{Talbot:2024yqw}, \texttt{gwforge}~\citep{Chandra:2024dhf}, \texttt{gwfish}~\citep{Dupletsa:2022scg}.
This document has LIGO preprint number {\tt LIGO-P2600254}.


\appendix
\section{Details of the star formation and merger rate models}
\label{app:fit_parameters}

\begin{figure*}
\centering
    \includegraphics[width=0.48\linewidth]{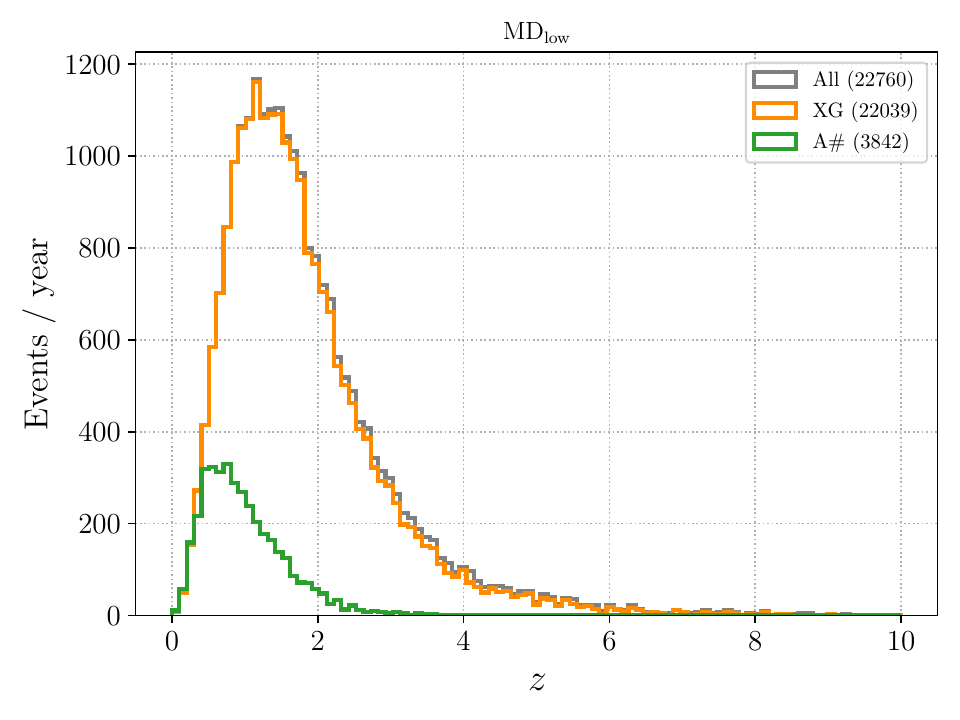}
    \includegraphics[width=0.48\linewidth]{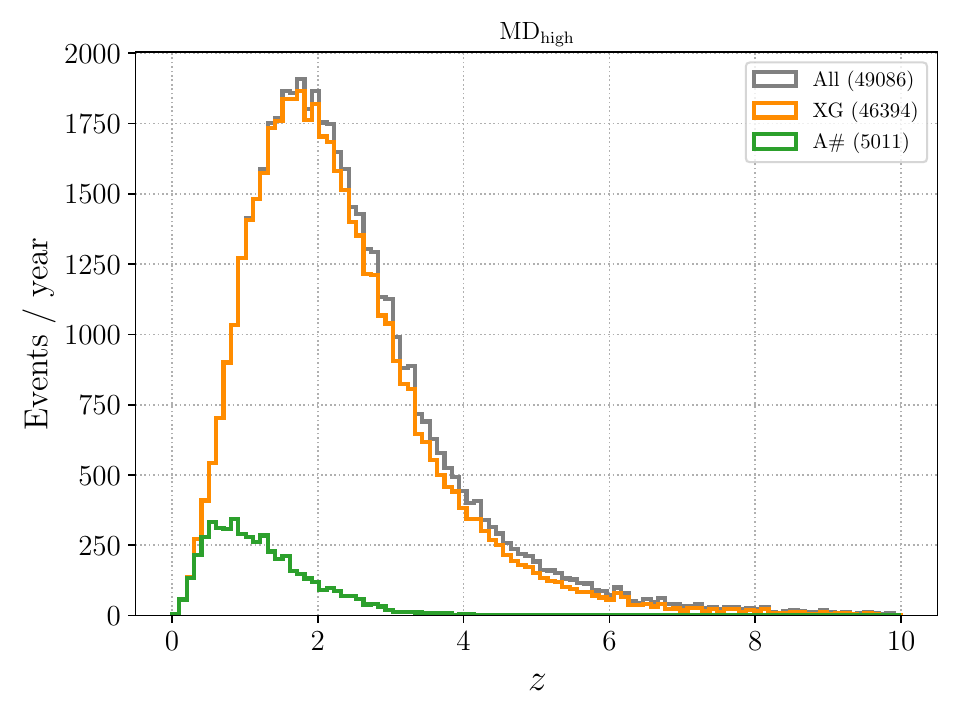}
    \caption{Input population for \mdhigh\ and \mdlow\ population models. The histograms in grey show all the simulated BBH systems for 1 year of data up to a redshift of 10. The detected events for XG and A\# networks have been plotted in orange and green histograms, respectively.}
    \label{fig:input_pop_MDhigh_MDlow}
\end{figure*}

In Sec.~\ref{subsec:method:setup}, we consider three different redshift evolutions for the star formation rate.  In Fig.~\ref{fig:input-pop-z}, we show the population of mergers for the MD model, and the observable signals in A\# and next-generation observatories.  For completeness, in Fig.~\ref{fig:input_pop_MDhigh_MDlow}, we show the population associated to the \mdhigh\ and \mdlow\ models and the observable signals. Table \ref{tab:hyperpars} gives the parameters for the population models for these three scenarios. 

\begin{figure}
    \centering
    \includegraphics[width=\linewidth]{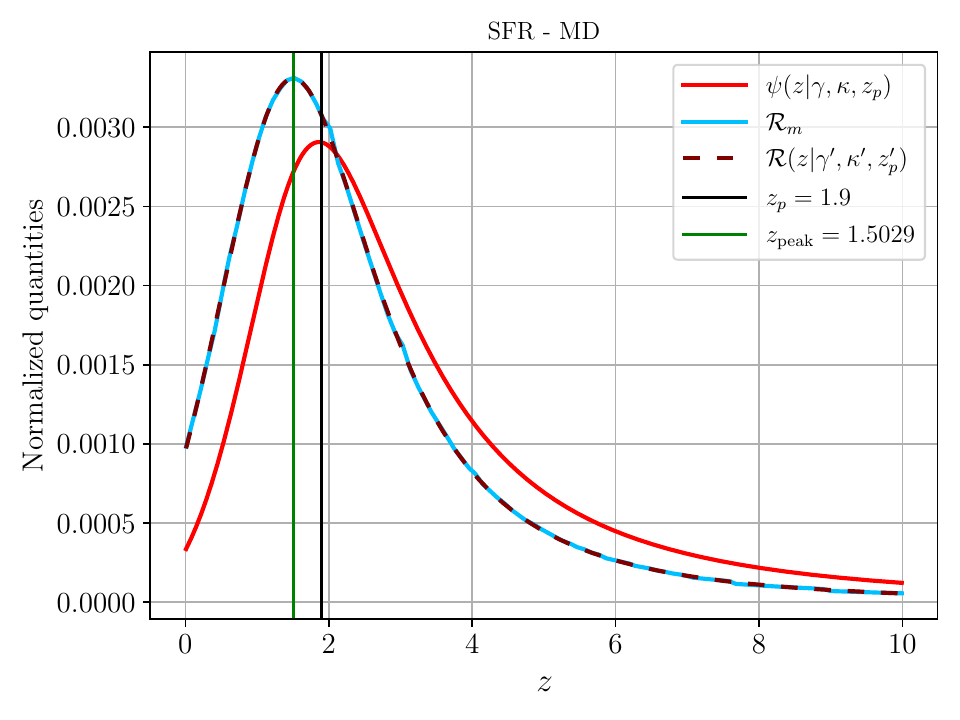}
    \caption{Plot showing that the merger rate $\mathcal{R}_m$ after the inclusion of time delay model can be parametrized by $\mathcal{R}(z|\gamma', \kappa', z_p')$ as shown in Eq.~\eqref{eq:R_z} for MD population model. Red curve is the Madau-Dickinson equation $\psi(z)$ before the inclusion of time delay whereas light blue curve shows the merger rate $\mathcal{R}_m$ after time delay is included. Dark maroon curve is $\psi(z)$ plotted using the modified set of parameters ($\gamma', \kappa', z_p'$) and matches quite well with the blue curve when normalized. We have also shown that the red curve peaks at $z_p$ and the dark maroon curve peaks at $z_\text{peak}$.}
    \label{fig:fitted-curve}
\end{figure}

Fig.~\ref{fig:fitted-curve} shows that the merger rate $\mathcal{R}_m$ after the inclusion of time delay model can be parametrized by $\mathcal{R}(z|\gamma', \kappa', z_p')$ as shown in Eq.~\eqref{eq:R_z} for MD population model. Red curve is the Madau-Dickinson equation $\psi(z)$ before the inclusion of time delay whereas light blue curve shows the merger rate $\mathcal{R}_m$ after time delay is included. Dark maroon curve is $\psi(z)$ plotted using the modified set of parameters ($\gamma', \kappa', z_p'$) and matches quite well with the blue curve when normalized. We have also shown that the red curve peaks at $z_p$ and the dark maroon curve peaks at $z_\text{peak}$. The values of modified parameters ($\gamma', \kappa', z_p'$) are given in Table \ref{tab:hyperpars}.

\begin{table}
\centering
\def\arraystretch{1.5}
\setlength{\tabcolsep}{8pt}
\begin{tabular}{|c|c|c|c|}
\hline
Parameter & \mdlow & MD & \mdhigh \\ \hline
$\gamma$ & 2.7 & 2.7 & 2.7 \\ \hline
$\kappa$ & 5.6 & 5.6 & 5.6 \\ \hline
$z_p$ & 1.54 & 1.9 & 2.53 \\ \hline
$\gamma'$ & 1.81 & 1.80 & 1.80 \\ \hline
$\kappa'$ & 5.28 & 5.30 & 5.32 \\ \hline
$z_p'$ & 1.49 & 1.84 & 2.42 \\ \hline
$z_\text{peak}$ & 1.20 & 1.50 & 2.01 \\ \hline
\end{tabular}
\caption{
Population parameter values from Eqs.~\eqref{eq:madau-dickinson}, \eqref{eq:R_z} and \eqref{eq:zp_tilde} for \mdlow, MD, and \mdhigh\ population models used for redshift.}
\label{tab:hyperpars}
\end{table}

\begin{table}
\centering
\def\arraystretch{1.4}
\setlength{\tabcolsep}{8pt}
\begin{tabular}{|c|c|}
\hline
\textbf{Parameter} & \textbf{Prior bounds} \\ \hline
$\gamma'$          & Uniform (0 - 5)       \\ \hline
$\kappa'$          & Uniform (0 - 20)      \\ \hline
$z_p'$   & Uniform (0.5 - 4)     \\ \hline
$z_\text{peak}$      & Constraint (0 - 5)    \\ \hline
\end{tabular}
\caption{Priors on the population parameters for redshift model used for population inference.}
\label{tab:priors}
\end{table}

\section{Detector locations}
\label{app:detectors}

In Table \ref{table:det-loc}, we provide the locations of the detectors used in this study.  For networks with A+ and A\# sensitivity, we place detectors at Hanford, Livingston and Aundha.  For the next-generation network we use two cosmic explorer observatories in the US and one triangular ET. Since the locations of the next generation detectors have not yet been decided, we use 
and place Einstein Telescope at the location of the Virgo detector, CE40 at the default location mentioned in the \texttt{LALSuite} package~\citep{lalsuite}, and CE20 off the coast of the US.

\begin{table}
\centering
\def\arraystretch{1.3}
\setlength{\tabcolsep}{7pt}
\begin{tabular}{|c|c|c|c|c|}
\hline
Label & Latitude      & Longitude       & Azimuth        & Network \\ \hline
L     & $30.56^\circ$ & $-90.77^\circ$  & $197.7^\circ$  & A\#, A+        \\ \hline
H     & $46.46^\circ$ & $-119.4^\circ$  & $126^\circ$    & A\#, A+        \\ \hline
I     & $19.61^\circ$ & $77.03^\circ$   & $117.62^\circ$ & A\#, A+        \\ \hline
CE40  & $34.17^\circ$ & $-118.13^\circ$ & $90^\circ$     & XG              \\ \hline
CE20  & $29^\circ$    & $-94^\circ$     & $-200^\circ$   & XG              \\ \hline
ET    & $43.63^\circ$ & $10.50^\circ$   & $19.43^\circ$  & XG              \\ \hline
\end{tabular}
\caption{The detector locations which have been used in this study. All angle values are given in degrees.}
\label{table:det-loc}
\end{table}


\bibliographystyle{mnras}
\bibliography{master_refs_new}
\label{lastpage}
\end{document}